\documentclass[twocolumn]{aastex63}

\usepackage{color}
\usepackage{CJK}
\usepackage{graphicx}
\usepackage{amsmath}
\usepackage{amssymb}
\usepackage{appendix}

\hypersetup{pdfauthor={Name}}

\graphicspath{{./}{figures/}}

\shorttitle{$R$-Process Enrichment}
\shortauthors{Nugent et al.}

\begin{document}

\title{Where has all the $r$-process gone? Timescales for GRB-Kilonovae to Enrich their Host Galaxies} 

\correspondingauthor{A. E. Nugent}
\email{anya.nugent@cfa.harvard.edu}

\newcommand{\NU}{\affiliation{Center for Interdisciplinary Exploration and Research in Astrophysics (CIERA) and Department of Physics and Astronomy, Northwestern University, Evanston, IL 60208, USA}}

\newcommand{\Purdue}{\affiliation{Purdue University, 
Department of Physics and Astronomy, 525 Northwestern Avenue, West Lafayette, IN 47907, USA}}

\newcommand{\CfA}{\affiliation{Center for Astrophysics\:$|$\:Harvard \& Smithsonian, 60 Garden St. Cambridge, MA 02138, USA}}

\newcommand{\UCSC}{\affiliation{Department of Astronomy and Astrophysics, University of California, Santa Cruz, CA 95064, USA}}

\newcommand{\IS}{\affiliation{Centre for Astrophysics and Cosmology, Science Institute, University of Iceland, Dunhagi 5, 107 Reykjav\'ik, Iceland}}

\newcommand{\DAWN}{\affiliation{Cosmic Dawn Center (DAWN), Niels Bohr Institute, University of Copenhagen, Jagtvej 128, 2100 Copenhagen \O, Denmark}}

\newcommand{\PUCV}{\affiliation{Instituto de F\'isica, Pontificia Universidad Cat\'olica de Valpara\'iso, Casilla 4059, Valpara\'iso, Chile}}

\newcommand{\IPMU}{\affiliation{Kavli Institute for the Physics and Mathematics of the Universe (Kavli IPMU), 5-1-5 Kashiwanoha, Kashiwa, 277-8583, Japan}}

\newcommand{\PSU}{\affiliation{Department of Astronomy \& Astrophysics, The Pennsylvania State University, University Park, PA 16802, USA}}

\newcommand{\ICDS}{\affiliation{Institute for Computational \& Data Sciences, The Pennsylvania State University, University Park, PA, USA}}

\newcommand{\IGC}{\affiliation{Institute for Gravitation and the Cosmos, The Pennsylvania State University, University Park, PA 16802, USA}}

\newcommand{\Swin}{\affiliation{ Centre for Astrophysics and Supercomputing, Swinburne University of Technology, Hawthorn, VIC, 3122, Australia}}

\newcommand{\Curtin}{\affiliation{ International Centre for Radio Astronomy Research, Curtin University, Bentley, WA 6102, Australia}}

\newcommand{\MQ}{\affiliation{Department of Physics \& Astronomy, Macquarie University, NSW 2109, Australia}}

\newcommand{\MQAAAstro}{\affiliation{Macquarie University Research Centre for Astronomy, Astrophysics \& Astrophotonics, Sydney, NSW 2109, Australia}}

\newcommand{\CSIRO}{\affiliation{CSIRO, Space and Astronomy, PO Box 76, Epping NSW 1710 Australia}}

\newcommand{\KICP}{\affiliation{Kavli Institute for Cosmological Physics, The University of Chicago, 5640 South Ellis Avenue, Chicago, IL 60637, USA}}

\newcommand{\UChicago}{\affiliation{Department of Astronomy \& Astrophysics, University of Chicago, 5640 S Ellis Avenue, Chicago, IL 60637, USA}}

\newcommand{\UA}{\affiliation{University of Arizona, Steward Observatory, 933~N.~Cherry~Ave., Tucson, AZ 85721, USA}}

\newcommand{\EFI}{\affiliation{Enrico Fermi Institute, The University of Chicago, 933 East 56th Street, Chicago, IL 60637, USA}}

\newcommand{\mpia}{\affiliation{Max-Planck-Institut f\"{u}r Astronomie (MPIA), K\"{o}nigstuhl 17, 69117 Heidelberg, Germany}}

\newcommand{\GWU}{\affiliation{Department of Physics, The George Washington University, Washington, DC 20052, USA}}

\newcommand{\UCB}{\affiliation{Department of Astronomy, University of California, Berkeley, CA 94720-3411, USA}}

\newcommand{\RU}{\affiliation{Department of Astrophysics/IMAPP, Radboud University, PO Box 9010,
6500 GL, The Netherlands}}

\newcommand{\LJMU}{\affiliation{Astrophysics Research Institute, Liverpool John Moores University, IC2, Liverpool Science Park, 146 Brownlow Hill, Liverpool L3 5RF, UK}}

\newcommand{\LU}{\affiliation{School of Physics and Astronomy, University of Leicester, University Road, Leicester. LE1 7RH, UK}}

\newcommand{\Adler}{\affiliation{The Adler Planetarium, 1300 South DuSable Lake Shore Drive, Chicago, IL 60605, USA}}

\newcommand{\ANU}{\affiliation{Research School of Astronomy and Astrophysics, Australian National University, Canberra, ACT 2611, Australia}}

\newcommand{\Car}{\affiliation{Cardiff Hub for Astrophysics Research and Technology, School of Physics \& Astronomy, Cardiff University, Queen's Buildings, Cardiff CF24 3AA, UK}}
\author[0000-0002-2028-9329]{Anya E. Nugent}
\CfA\NU

\author[0000-0002-4863-8842]{Alexander P. Ji}
\UChicago
\KICP

\author[0000-0002-7374-935X]{Wen-fai Fong}
\NU

\author[0000-0002-9136-6731]{Hilay Shah}
\ANU 

\author[0000-0002-6301-638X]{Freeke van de Voort}
\Car

\begin{abstract}
Neutron star (NS) mergers are currently the only observed source of $r$-process production in the Universe. Yet, it is unclear how much $r$-process mass from these mergers is incorporated into star-forming gas to enrich stars. This is crucial to consider as all other $r$-process mass estimates in the Universe beyond Earth are based on stellar $r$-process abundances. Here, we explore the extent to which merger location and host galaxy properties affect the incorporation of $r$-process elements into star-forming gas, and quantify an ``enrichment" timescale to account for this process. To put this timescale in context, we analyze a population of 12 gamma-ray bursts (GRBs) with probable associations to $r$-process kilonovae (GRB-KNe) and 74 short GRBs without claimed KNe, including new non-parametric star formation histories for the GRB-KN hosts. We find enrichment timescales for this sample are between $\approx7$~Myr$-1.6$~Gyr, suggesting that environmental enrichment is delayed from NS merger occurrence. Moreover, we find a correlation between the amount of environmental enrichment from a single event and increasing host specific star formation rate (sSFR), and little correlation with stellar mass and GRB galactocentric offset. Environments with low sSFRs ($<10^{-10.5}$ yr$^{-1}$), which comprise 18\% of short GRB hosts and the host of GW170817, will have little to no capacity for stellar enrichment. Our results indicate that not all $r$-process from NS mergers is incorporated into newly-forming stars, and instead some remains ``lost" to the CGM or IGM. Future studies should consider these losses to understand the total contribution from NS mergers to the Universe's $r$-process budget.
\end{abstract}


\section{Introduction}
\label{sec:intro}

The astrophysical birthsites of heavy $r$-process elements ($A > 130$) have long been debated. There are several proposed channels of $r$-process, including neutron star (NS) mergers \citep{eichler1989, ber14, bhp2016, kks+2016, hbp+2018} and rare types of core-collapse supernovae (CCSNe; \citealt{quian2020, astq2004, tisa2015, hm2018, sbm2019, shsc2019, Brauer2021}). While NS mergers are subject to a wide range of progenitor formation timescales (delay times) and host environments \citep{bkd+2010, dbk+2012, obb+2017}, collapse of massive stars (collapsars) and CCSNe are almost exclusively connected to short delay times and star-forming host galaxies \citep{Svensson2010, Perley2013, Vergani2015, Wang2014, bbf+17, Niino2017, Schulze2021, tp2021}. Given this diversity in progenitor timescales and environments, the channel(s) through which $r$-process material is synthesized has substantial impact on the chemical enrichment of the Universe, as well as how our solar system achieved its $r$-process mass and abundance pattern.

Despite relevant observations of a handful of CCSNe and collapsars to determine if they produce any $r$-process, there are no clear signs of $r$-process element production in these events thus far \citep{bvc+2023, aby+2024, rfl+2024}. On the other hand, we have direct evidence of $r$-process production in NS mergers, with the coincident detection of an $r$-process kilonova (KN) with a binary NS merger, gravitational wave (GW) event GW170817 \citep{aaloc+17, gw170817mma, cbk+17, kmb+17, mhh+17, pdb+17, ssd+2017, tlg+17}.  Because of the coincidence of GW170817 to short gamma-ray burst (GRB) 170817A \citep{aaloc+17, gvb+17, sfk+2017}, it is also now confirmed that at least some short GRBs are connected to the local population of GW-NS mergers and contribute to $r$-process enrichment. Additionally, a number of short GRBs have been followed by photometric near-infrared excesses, presumed to be KNe \citep{bfc+2013, tlf+2013, jlc+2015, yjl+2015, fmc+2016, jhl+2016, kkl+2017, glt+2018, trp+2018, gtl+2019, jcl+2020, rsm+2020, flr+2021, otd+2021, rkp+2021}. 

Observations of $r$-process elements across a wide variety of environments have further supported an NS merger production pathway. The stochastic $r$-process abundance pattern observed in Galactic metal poor stars \citep{elp+89, McWilliam1995,cfb+2018} and low metallicity Local Group dwarf galaxies \citep{Shetrone2001,Venn2012,jfc+2016, mht+2021, mrr+2021,Reggiani2021,njc+2022,Limberg2023} suggests that $r$-process events are much rarer than normal CCSNe. Indeed, the $r$-process event rate and mass yield estimates from inferred stellar $r$-process abundances are well-matched to those derived for GW170817 and short GRBs \citep{hbp+2018, mr+2018, jfc+2016, rfb+2023}. This implies that NS mergers may be responsible for the majority of $r$-process elements in the Universe.

To complicate the picture, three long GRBs, GRBs 060614 \citep{jlc+2015,  yjl+2015, glt+2018, rsm+2020, rkp+2021}, 211211A \citep{rgl+2022, tfo+2022, yaz+2022} and 230307A \citep{gtf+2023, lgs+2024, yto+2024}, which typically derive from collapsars, had photometric excesses consistent with KNe. The potential association of these events to KNe not only implies they have merger (as opposed to collapsar) progenitors, but also that long GRBs can also contribute to the Universe's $r$-process budget. Taken at face value, the addition of this long GRB $r$-process channel may suggest that the total mass yield of $r$-process material from NS mergers exceeds the mass observed within Galactic metal poor stars and Local Group dwarfs \citep{clc+2024}. This dilemma only becomes more serious if rare types of CCSNe also contribute appreciably to the $r$-process budget.

Given that $r$-process studies of Galactic metal poor stars and Local Group dwarf galaxies focus on the $r$-process enrichment of stars, it is imperative to consider how $r$-process material from various production channels is incorporated into star forming gas to understand their potential contribution in these environments. The delay between $r$-process events and the redistribution of the metals into star-forming gas, or the ``enrichment timescale", may be comparable to progenitor delay times. Indeed, if this enrichment timescale is comparatively very long, the $r$-process mass may not be able to enrich star-forming gas and would remain in the circumgalactic medium (CGM) or intergalactic medium (IGM). This would result in ``losses'' when we compare the $r$-process material produced to that which makes it into stars. For NS mergers, their older stellar populations and natal kicks could result in substantial losses. For instance, GW170817 occurred in an old, massive host with little ongoing star formation \citep{bbf+17, llt+17, pht+17, kfb+2022}, and, thus, likely a small amount of star-forming gas left to enrich at the time of merger. Beyond GW170817, $\approx 15\%$ of short GRBs are associated with transitioning and quiescent host galaxies that have low ongoing star-formation rates of $\approx 0.14$~$M_\odot$~yr$^{-1}$ \citep{BRIGHT-II, ji2024}. Furthermore, NS mergers in general occur on more delayed timescales than, for example, collapsar events following a star-forming burst within their hosts \citep{Nakar2006, bfp+07, Jeong2010, Hao2013, Wanderman2015, tkf+17, Anand2018, am19, Zevin+DTD}. It therefore remains unclear what fraction of NS mergers are capable of enriching star-forming gas before the next possible star-forming period within their hosts.

In addition, neutron stars can experience significant natal kicks at their formation, translating to large systemic velocities, as observed by the Galactic binary NS population \citep{tkf+17, vns+18, az19}. Coupled with the long expected delay times of NS mergers, NS systems can easily migrate and merge well outside of their host galaxies \citep{zkn+2019, mlt+21, glc+2024}. This is backed by observations of the short GRB population, which have large galactocentric offsets of $\approx 5.6-7.7$~kpc \citep{cld+11, fb13, tlt+14, BRIGHT-I, otd+2022}. Subsequently, offsets may also play an important role in contributing to a prolonged $r$-process enrichment timescale. 

Here, we assess how significant $r$-process losses are from a population of 12 GRBs with claimed KNe (GRB-KNe). We model the GRB-KNe host galaxy stellar masses and star formation histories and use their galactocentric offsets to constrain the enrichment timescale and the fraction of stellar mass capable of enrichment from these events. We additionally compare these results to those determined for the literature sample of short GRBs without claimed KNe. In Section \ref{sec:sample}, we discuss our GRB sample. In Section \ref{sec:sp_model}, we describe the stellar population modeling methods used to determine host galaxy stellar masses and star formation histories and present our results on these and the hosts halo properties. We describe our method for quantifying the $r$-process enrichment timescale in Section \ref{sec:timescale}. We discuss major implications from these findings in Section \ref{sec:discussion}. Finally, we list our conclusions in Section \ref{sec:conclusion}.

Unless otherwise stated, all observations are reported in the AB magnitude system and have been corrected for Galactic extinction in the direction of the GRB \citep{MilkyWay,sf11}.  We employ a standard WMAP9 cosmology of $H_{0}$ = 69.6~km~s$^{-1}$~Mpc$^{-1}$, $\Omega_\textrm{m}$ = 0.286, $\Omega_\textrm{vac}$ = 0.714 \citep{Hinshaw2013, blw+14}.

\section{Sample}
\label{sec:sample}
\begin{deluxetable*}{l|ccccccccc}
\tabletypesize{\footnotesize}
\tablecolumns{10}
\tablewidth{0pc}
\tablecaption{GRB, Host, and Halo Properties
\label{tab:prop}}
\tablehead{
\colhead{GRB} &
\colhead{Sample} &
\colhead{z} &
\colhead{Projected Offset} &
\colhead{log($M_*/M_\odot$)} &
\colhead{SFR} &
\colhead{log($M_h/M_\odot$)} &
\colhead{$r_\textrm{vir}$} &
\colhead{$V_\textrm{vir}$} &
\colhead{$\log(T_\textrm{vir})$} \\
\colhead{} &
\colhead{} &
\colhead{} &
\colhead{[kpc]} &
\colhead{} &
\colhead{[$M_\odot$ yr$^{-1}$]} &
\colhead{} &
\colhead{[kpc]} &
\colhead{[km/s]} &
\colhead{[K]}
}
\startdata
050709 &  Gold & 0.161 & 3.76 & $9.07^{+0.05}_{-0.09}$ & $0.11^{+0.03}_{-0.02}$ & 11.31 & 152.83 & 75.6 & 5.31  \\ 
050724 & Gold & 0.257 & 2.74 &  $11.12^{+0.02}_{-0.01}$ & $0.2^{+0.02}_{-0.02}$ & 13.28 & 697.86 & 345.21 & 6.63  \\
060614 & Gold & 0.125 & 0.7 &  $7.77^{+0.11}_{-0.09}$ & $0.07^{+0.02}_{-0.02}$ & 10.66 & 93.46 & 46.23 & 4.88 \\ 
070714 & Gold & 0.925 & 12.33 &  $9.70^{+0.07}_{-0.09}$ & $1.89^{+0.89}_{-0.62}$ & 11.68 & 203.40 & 100.62 & 5.55  \\ 
070809  & Gold & 0.473 & 34.11 &  $10.9^{+0.14}_{-0.07}$ & $9.75^{+12.89}_{-8.55}$ & 12.97 & 546.21 & 270.20 & 6.41  \\ 
130603B  & Gold & 0.357 & 5.4 &  $9.66^{+0.15}_{-0.12}$ & $18.27^{+4.6}_{-4.89}$ & 11.59 & 189.48 & 93.73 & 5.50  \\ 
150101B  & Gold & 0.134 & 11.31  & $11.31^{+0.02}_{-0.02}$ & $1.84^{+0.57}_{-0.45}$ & 13.60 & 887.92 & 439.23 & 6.84  \\ 
160821B  & Silver & 0.162 & 15.74 & $9.44^{+0.03}_{-0.04}$ & $0.01^{+0.0}_{-0.0}$ & 11.48 & 175.07 & 86.60 & 5.43  \\
170817 & Gold & 0.0097 & 2.125 &  $10.80^{+0.04}_{-0.07}$ & $0.01^{+0.02}_{-0.01}$
 & 12.51 & 384.54 & 190.35 & 6.11  \\ 
200522A & Gold & 0.554 & 0.93 &  $9.50^{+0.04}_{-0.03}$ & $11.75^{+1.59}_{-1.84}$ & 11.53 & 181.54 & 89.81 & 5.46  \\
211211A & Gold & 0.076 & 7.92 &  $8.91^{+0.06}_{-0.06}$ & $0.35^{+0.04}_{-0.04}$ & 11.21 & 142.14 & 70.31 & 5.25  \\ 
230307A & Silver & 0.065 & 38.9 &  $9.66^{+0.09}_{-0.08}$ & $0.09^{+0.09}_{-0.06}$ & 11.60 & 190.7 & 94.34 & 5.50  \\ 
\enddata
\tablecomments{The sample (confidence of host association), spectroscopic redshifts, projected galactocentric offsets, stellar masses, present-day SFRs, halo masses, virial radii, virial velocities, and virial temperatures for our GRB sample. For stellar masses and SFRs, we report the median and 68\% confidence interval, as determined in our \texttt{Prospector} fits.}
\end{deluxetable*}

\subsection{GRB-Kilonova Sample}
Our main sample consists of 12 GRBs with probable KNe, confident host galaxy associations, and confirmed spectroscopic redshifts from their hosts. We focus on the sample with claimed KNe as it is reasonable to assume they produced $r$-process material. We begin with eight short GRBs that have claimed KNe, extensively discussed in the literature: GRBs 050709 \citep{jhl+2016, glt+2018, rsm+2020, rkp+2021}, 050724 \citep{rsm+2020, rkp+2021}, 070714B \citep{rsm+2020, rkp+2021}, 070809 \citep{jcl+2020, rsm+2020, rkp+2021}, 130603B \citep{bfc+2013, tlf+2013, glt+2018, rsm+2020, rkp+2021}, 150101B \citep{fmc+2016, trp+2018, rsm+2020, rkp+2021}, 160821B \citep{kkl+2017, gtl+2019, rsm+2020, rkp+2021}, and 200522A \citep{flr+2021, otd+2021, rkp+2021}. We further include GW170817/GRB 170817A, a known NS merger with spectroscopic evidence for a KN (AT 2017gfo; \citealt{cbk+17, kmb+17, mhh+17, pdb+17, ssd+2017, tlg+17}) that produced $\approx 0.05-0.08$~$M_\odot$ of $r$-process material \citep{vgb+2017, hbp+2018, rsf+2018}. 

Finally, we include three long GRBs that were followed by possible KNe: GRBs 060614 \citep{jlc+2015,  yjl+2015, glt+2018, rsm+2020, rkp+2021}, 211211A \citep{rgl+2022, tfo+2022, yaz+2022}, and the recently-detected long GRB 230307A \citep{gtf+2023, lgs+2024, yto+2024}. Relative to traditional Type Ic SNe following collapsars, all three GRBs had near-IR photometric excesses that peaked at earlier times, were less luminous, and faded more rapidly, suggesting these were not typical long GRB events. The lightcurves of the excesses for GRBs 211211A and 230307A also faded at rates similar to that of AT 2017gfo \citep{rgl+2022, lgs+2024}. These characteristics suggested that the observed excesses were KNe, and therefore that these events were merger-driven rather than deriving from collapsars. In addition, spectroscopic follow-up of the counterpart observed for GRB 230307A revealed detection of an emission feature at $\sim 2.15\,\mu$m. This was claimed to be Te~III at $z = 0.065$ \citep{gtf+2023, lgs+2024} although the identification is not definitive. A broad emission feature was observed for AT 2017gfo at a similar (but not exact) rest-frame wavelength, also proposed to be Te~III \citep{htkg2023}. If the identification is correct, this spectroscopically validates that the GRB 230307A derived from an NS merger and that the excess was an $r$-process KN.

\begin{figure*}[t]
\centering
\includegraphics[width=0.3\textwidth]{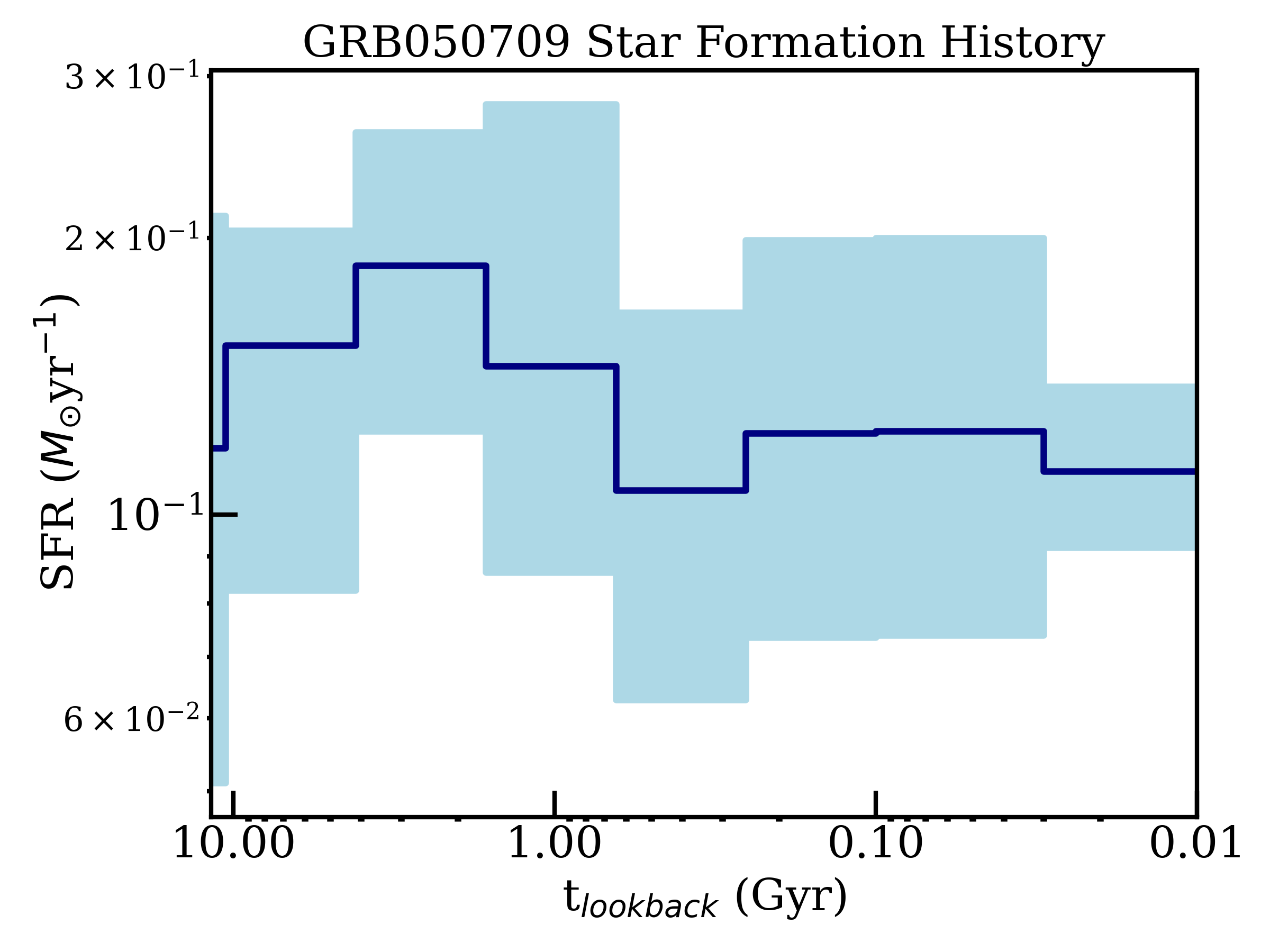}
\includegraphics[width=0.3\textwidth]{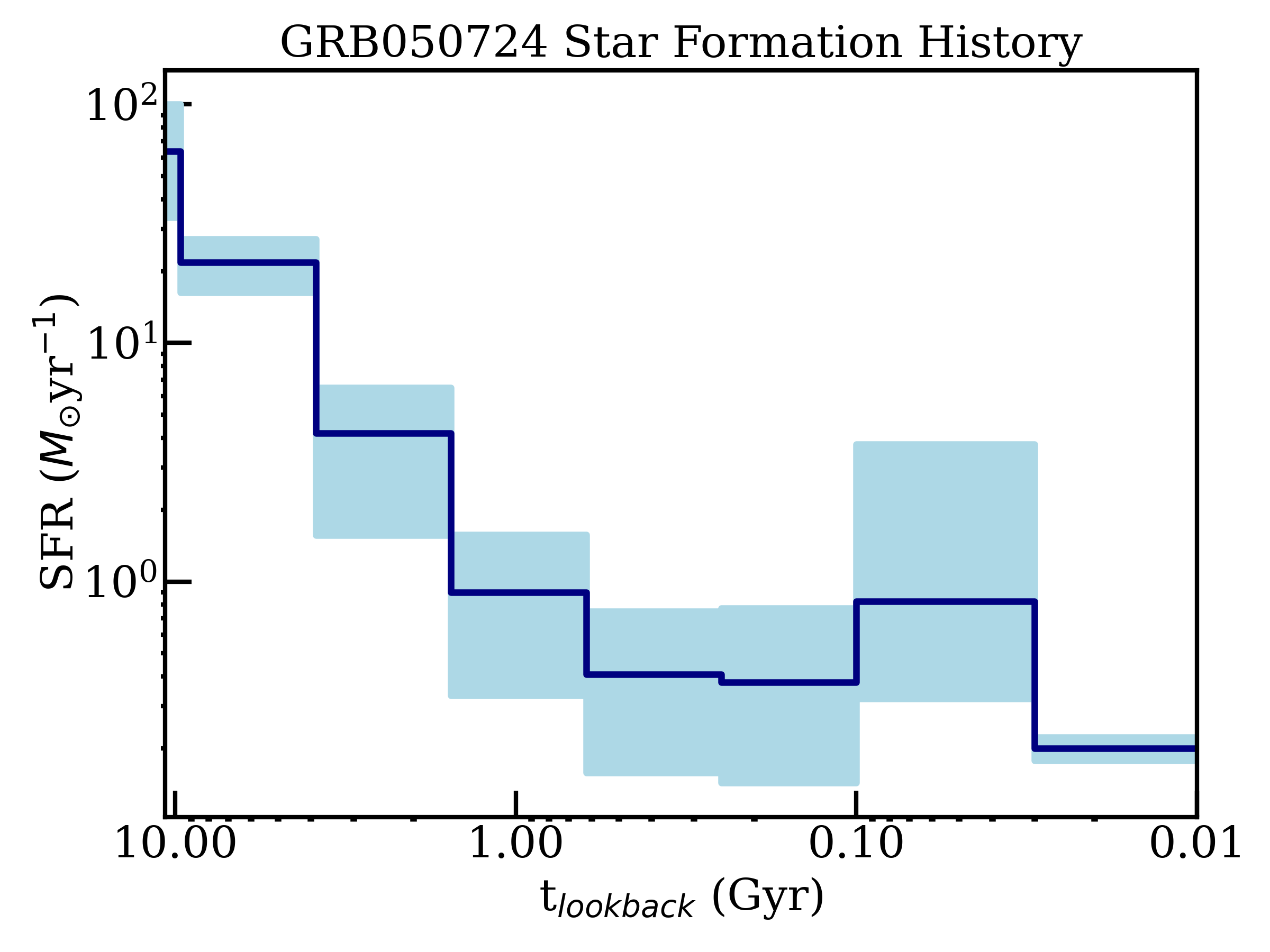}
\includegraphics[width=0.3\textwidth]{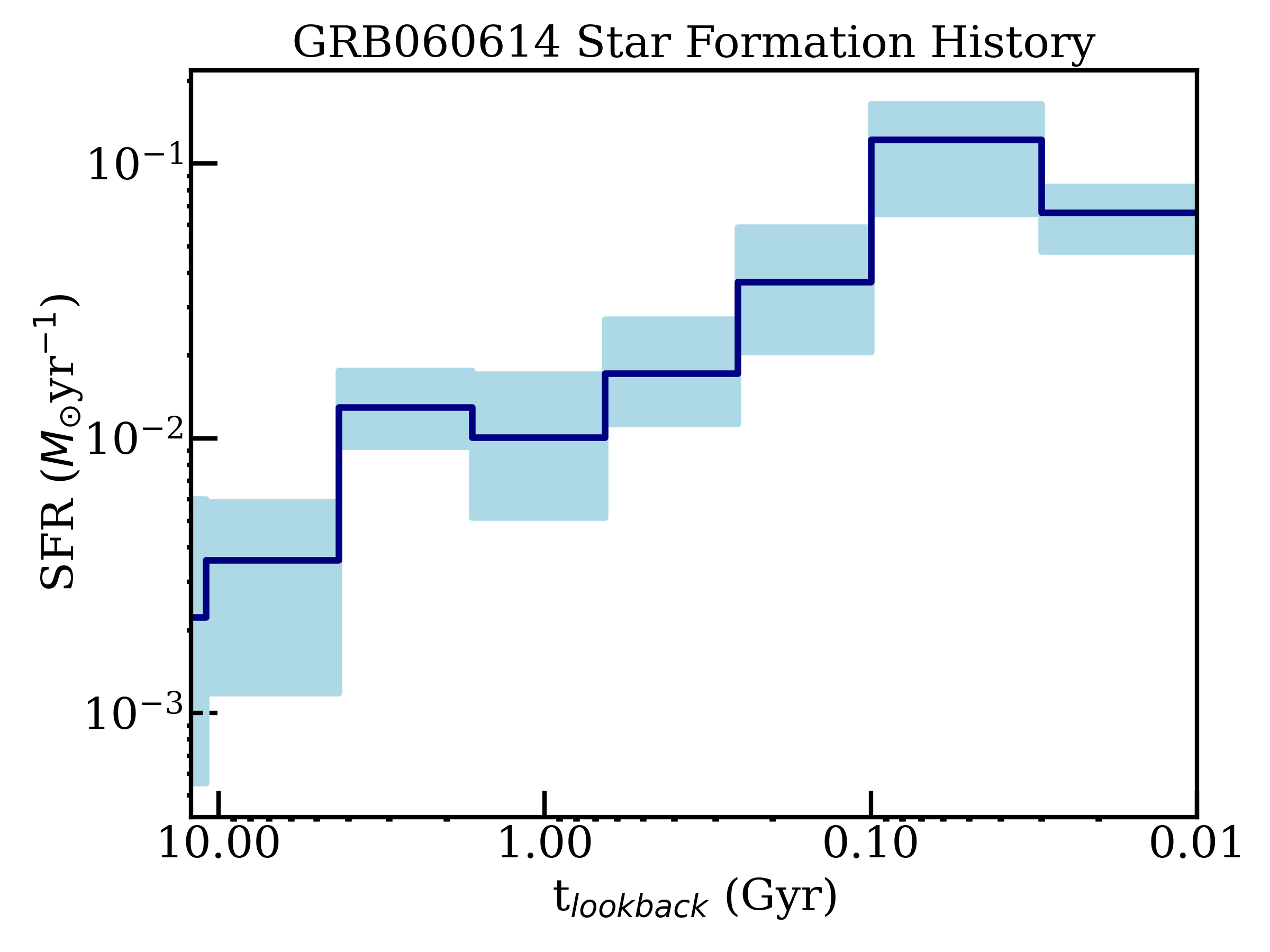}
\includegraphics[width=0.3\textwidth]{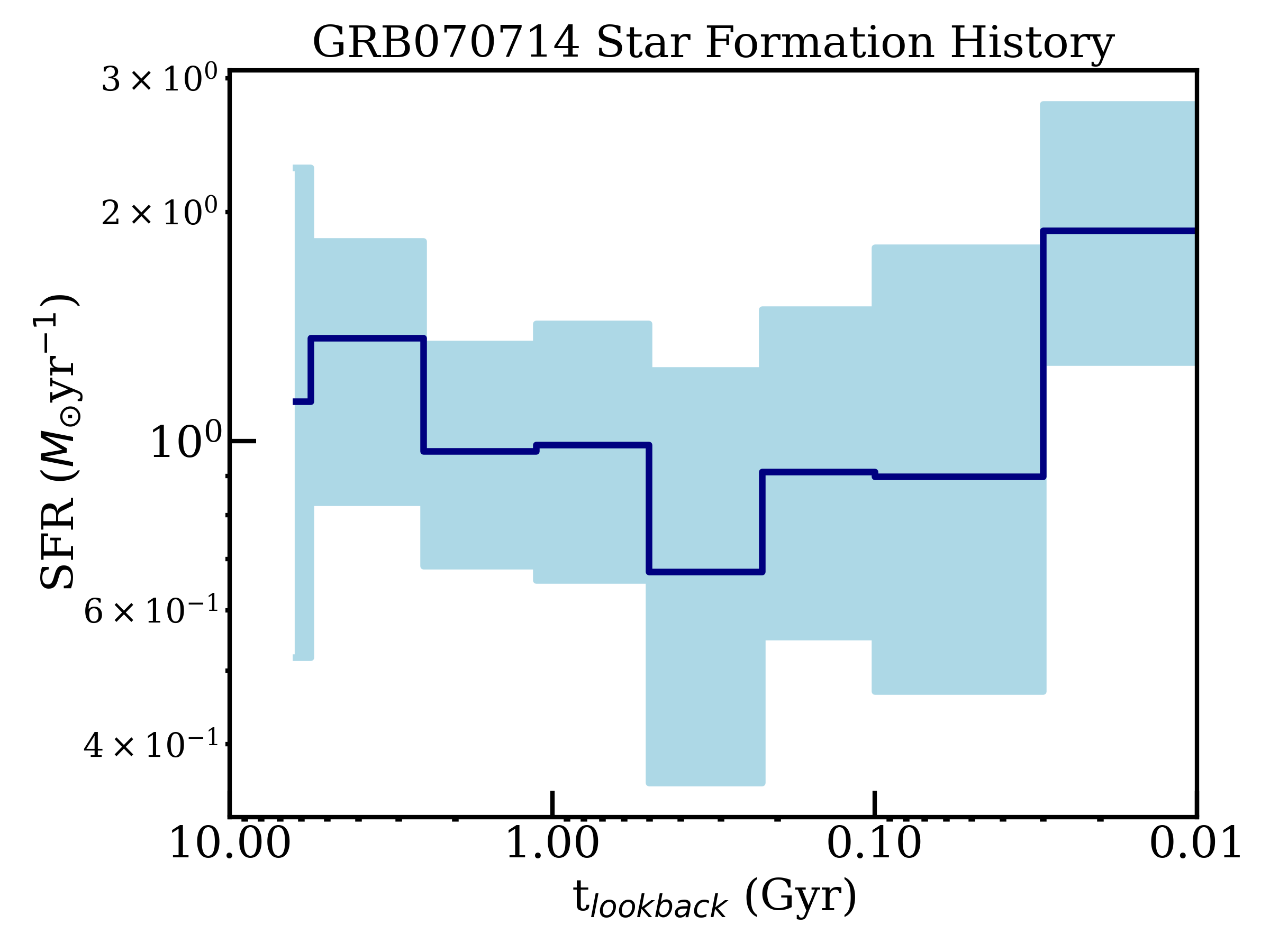}
\includegraphics[width=0.3\textwidth]{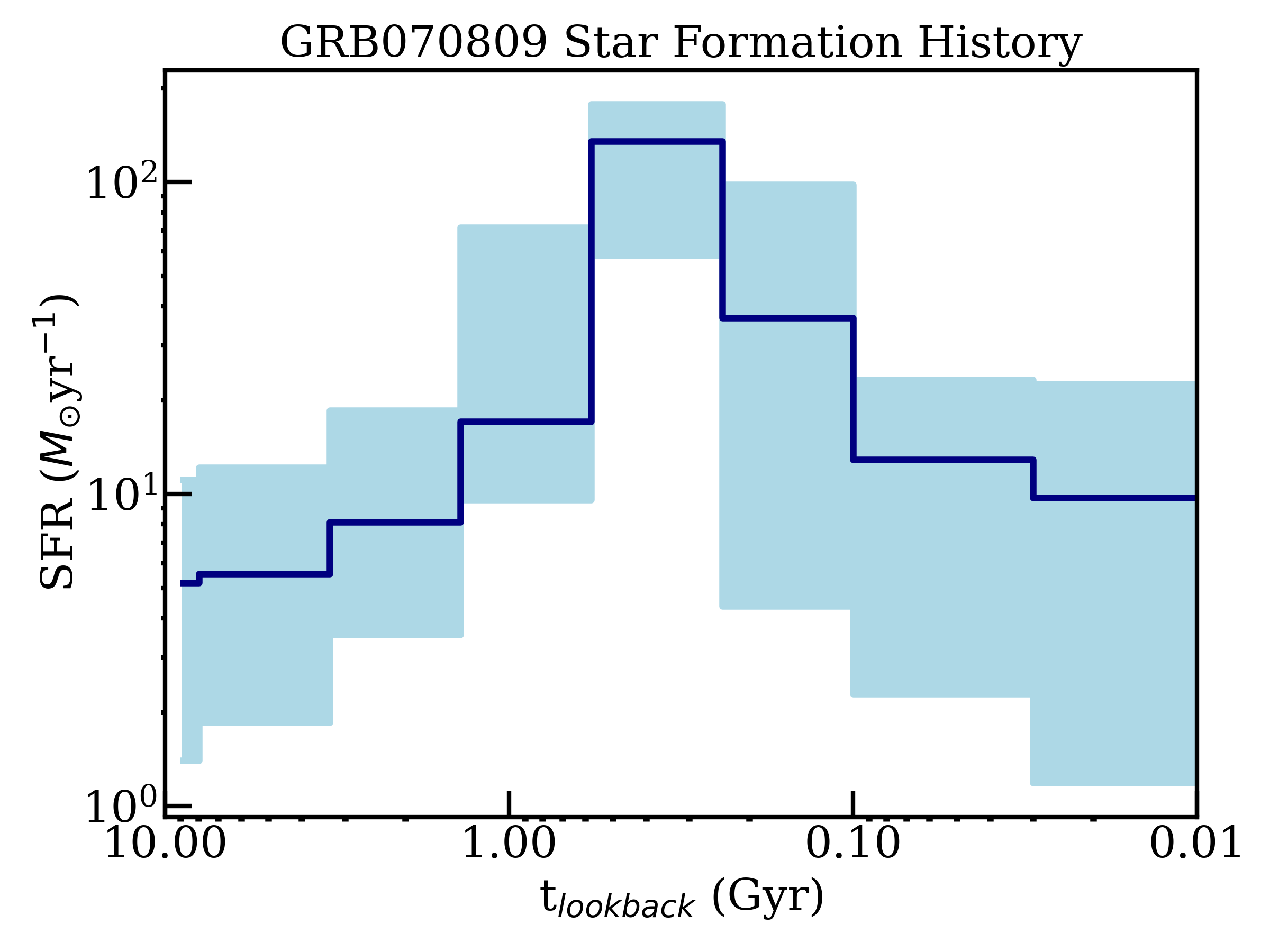}
\includegraphics[width=0.3\textwidth]{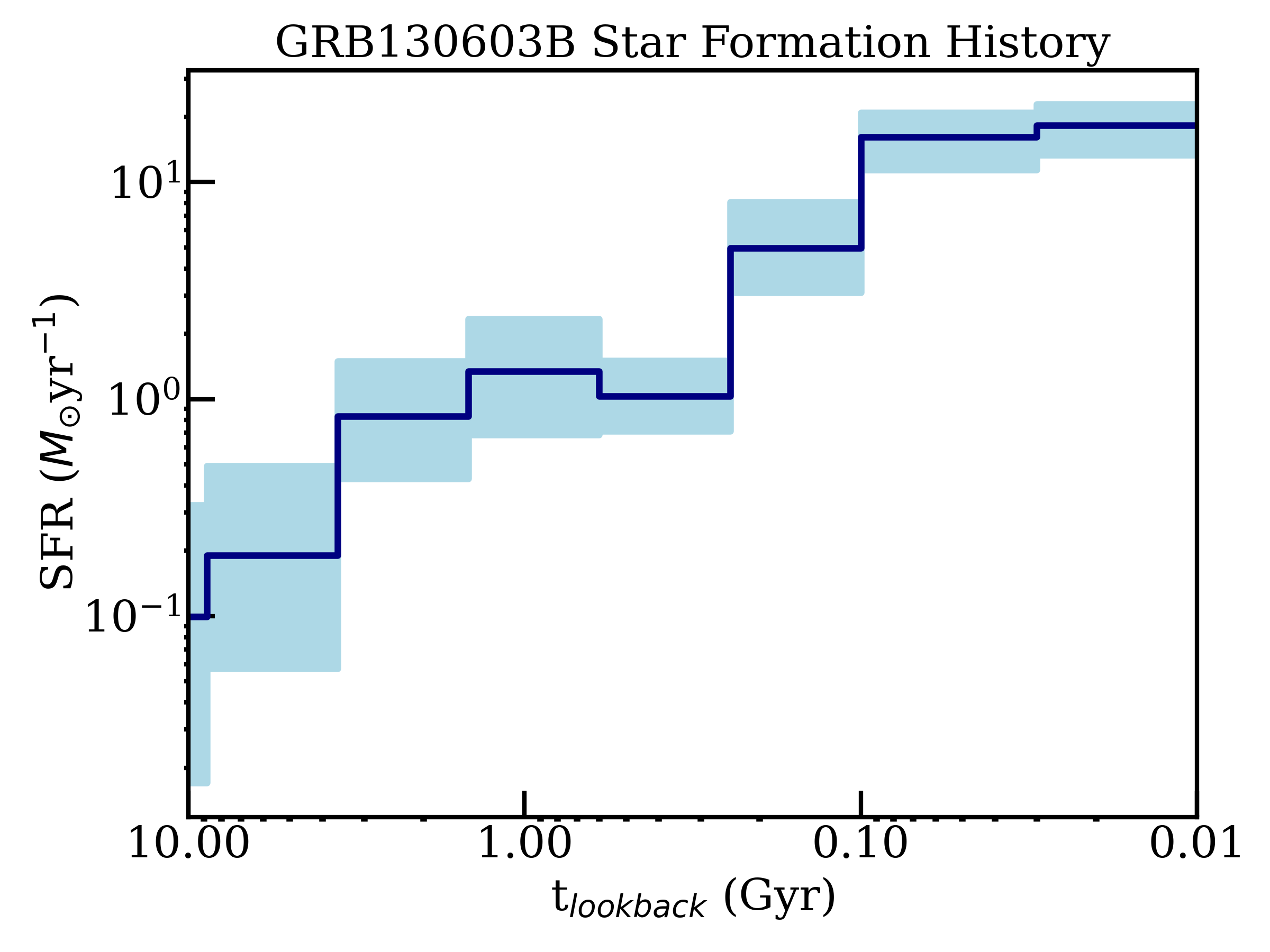}
\includegraphics[width=0.3\textwidth]{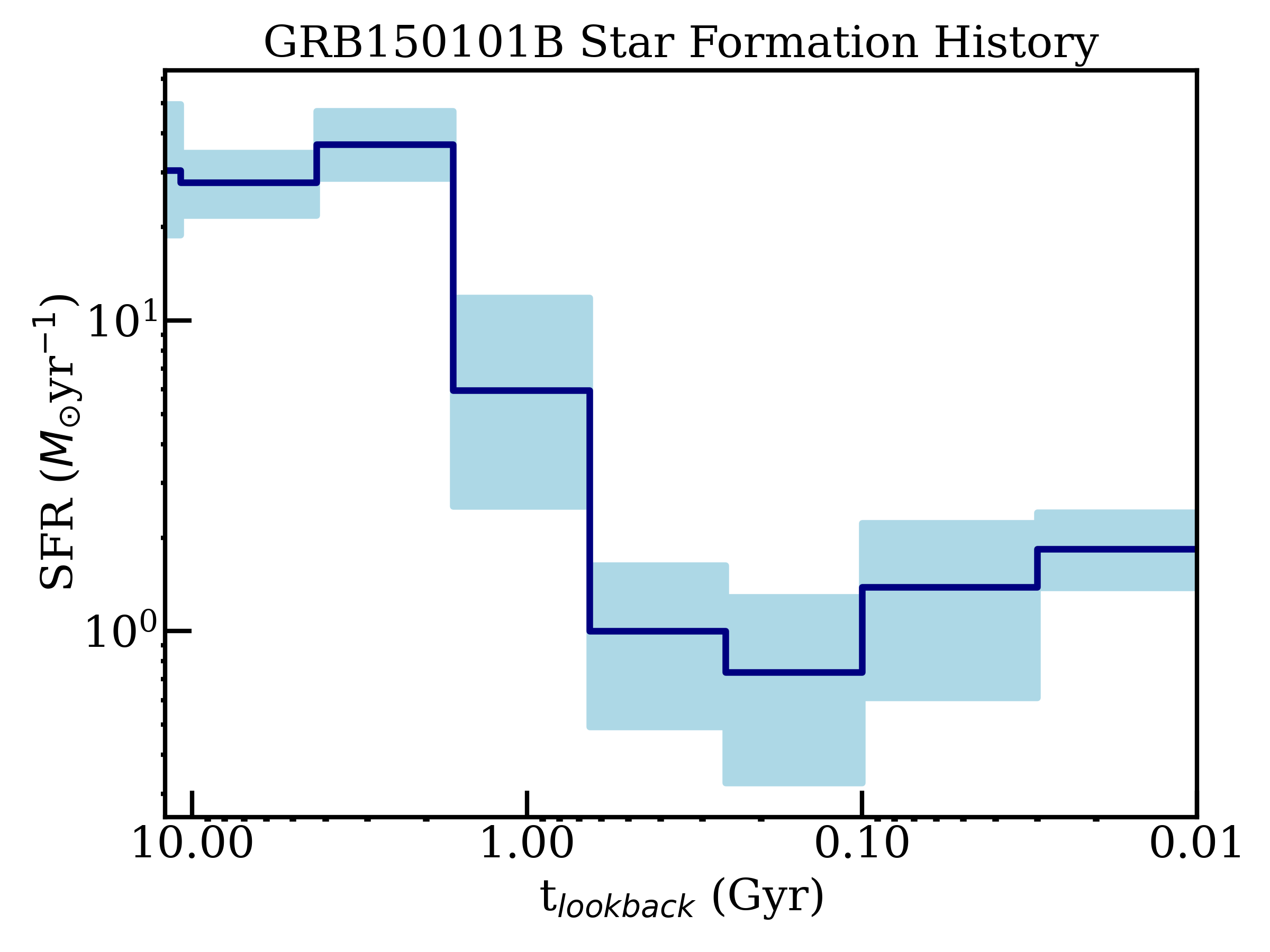}
\includegraphics[width=0.3\textwidth]{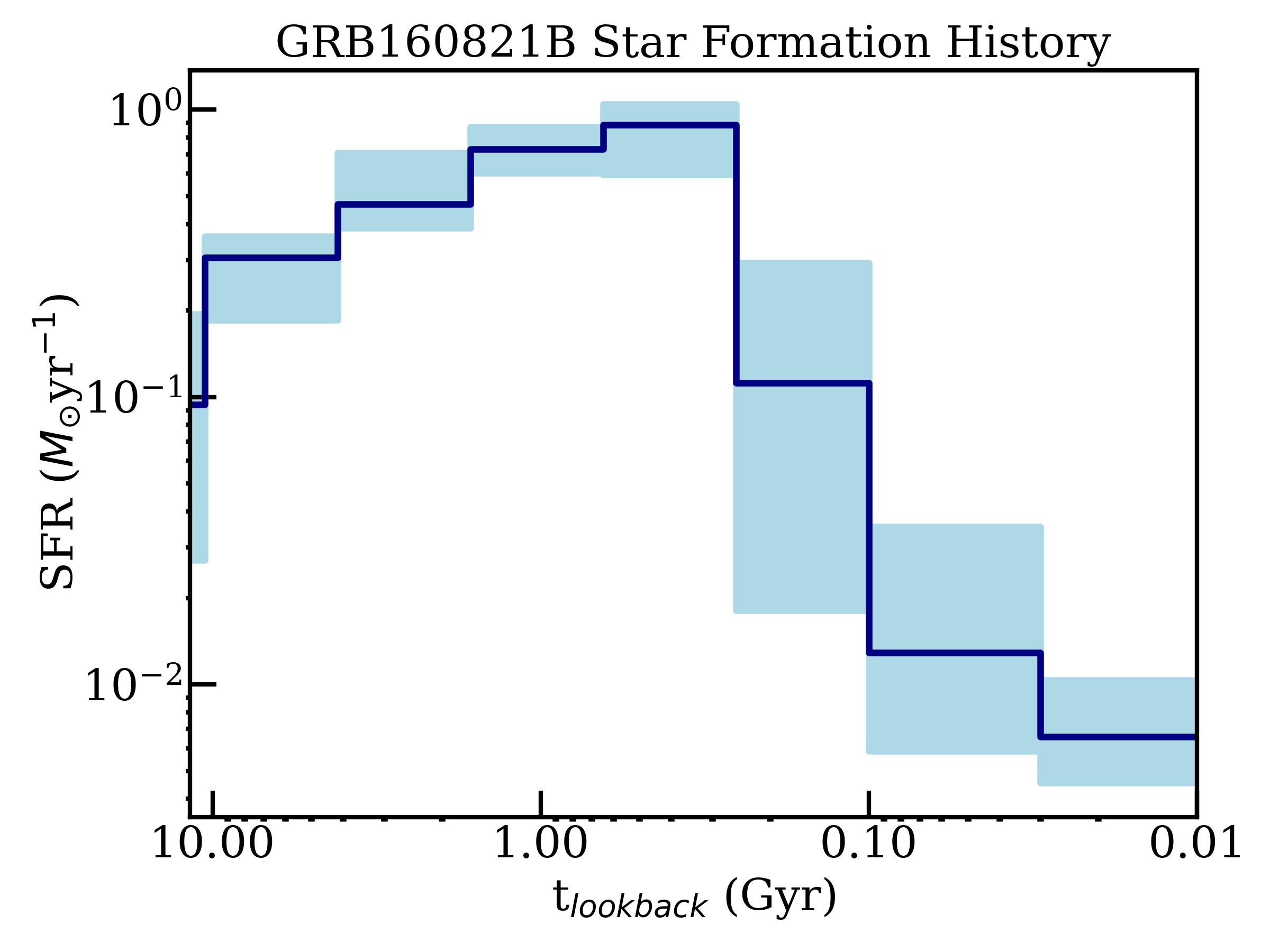}
\includegraphics[width=0.3\textwidth]{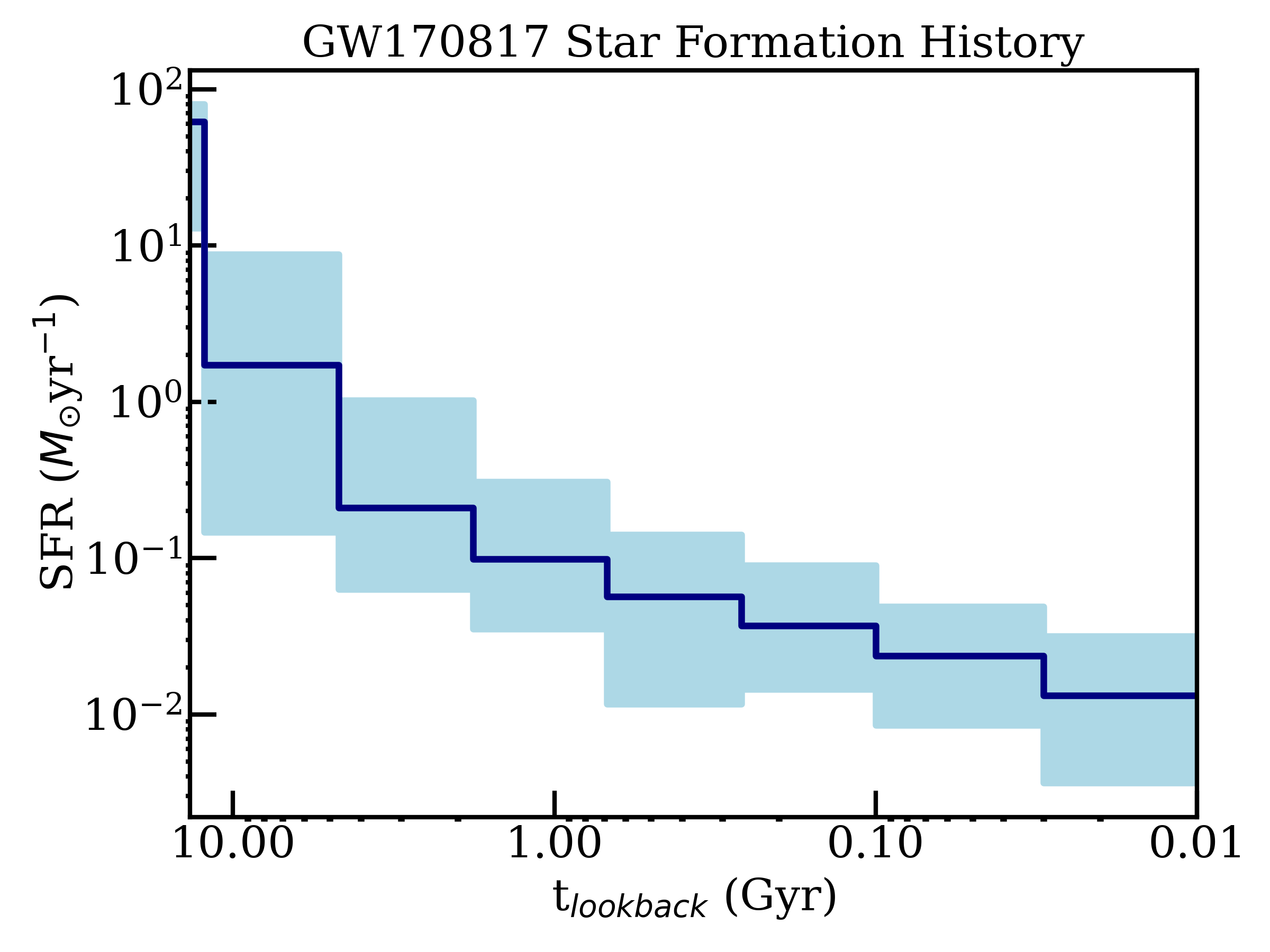}
\includegraphics[width=0.3\textwidth]{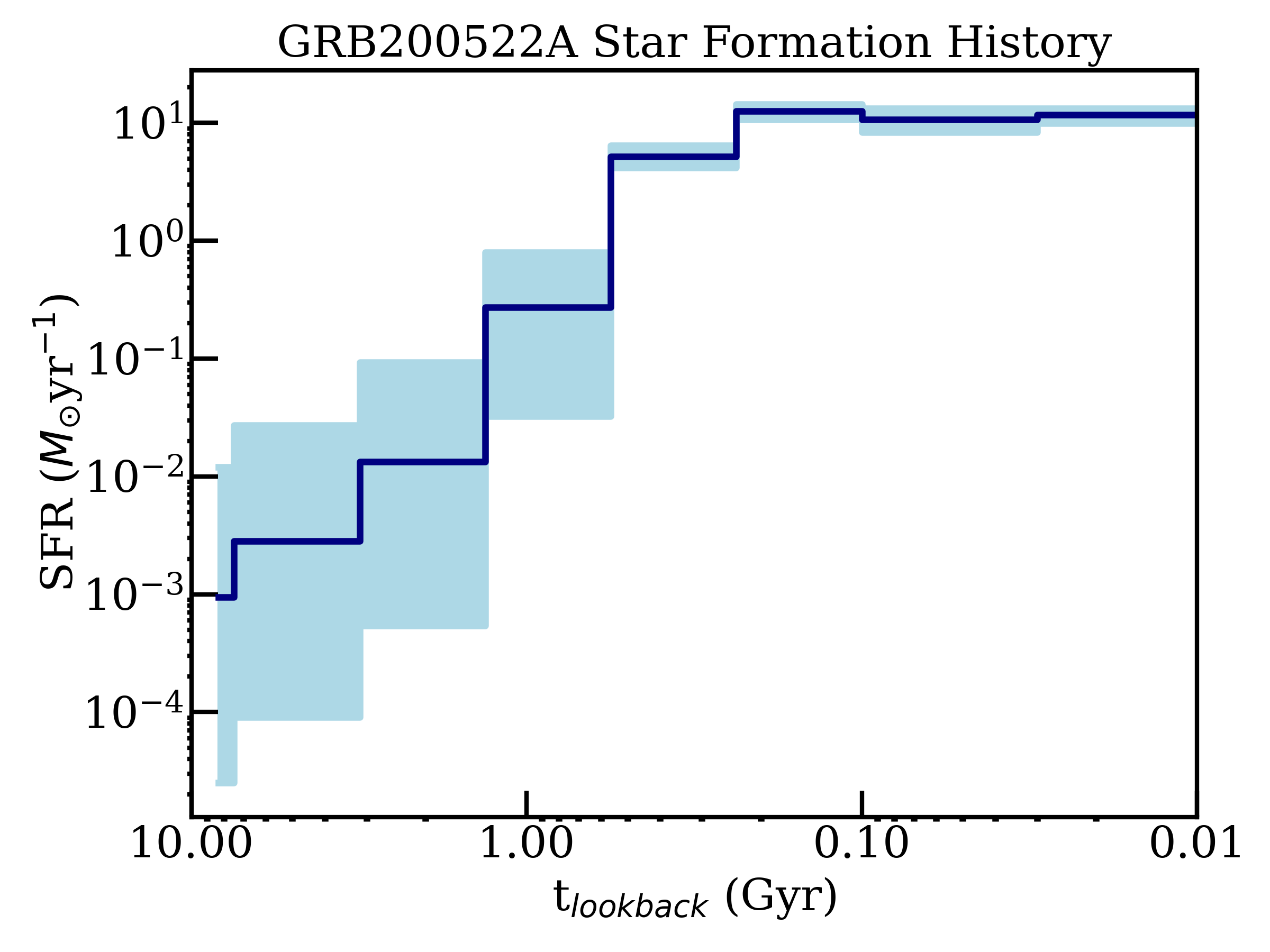}
\includegraphics[width=0.3\textwidth]{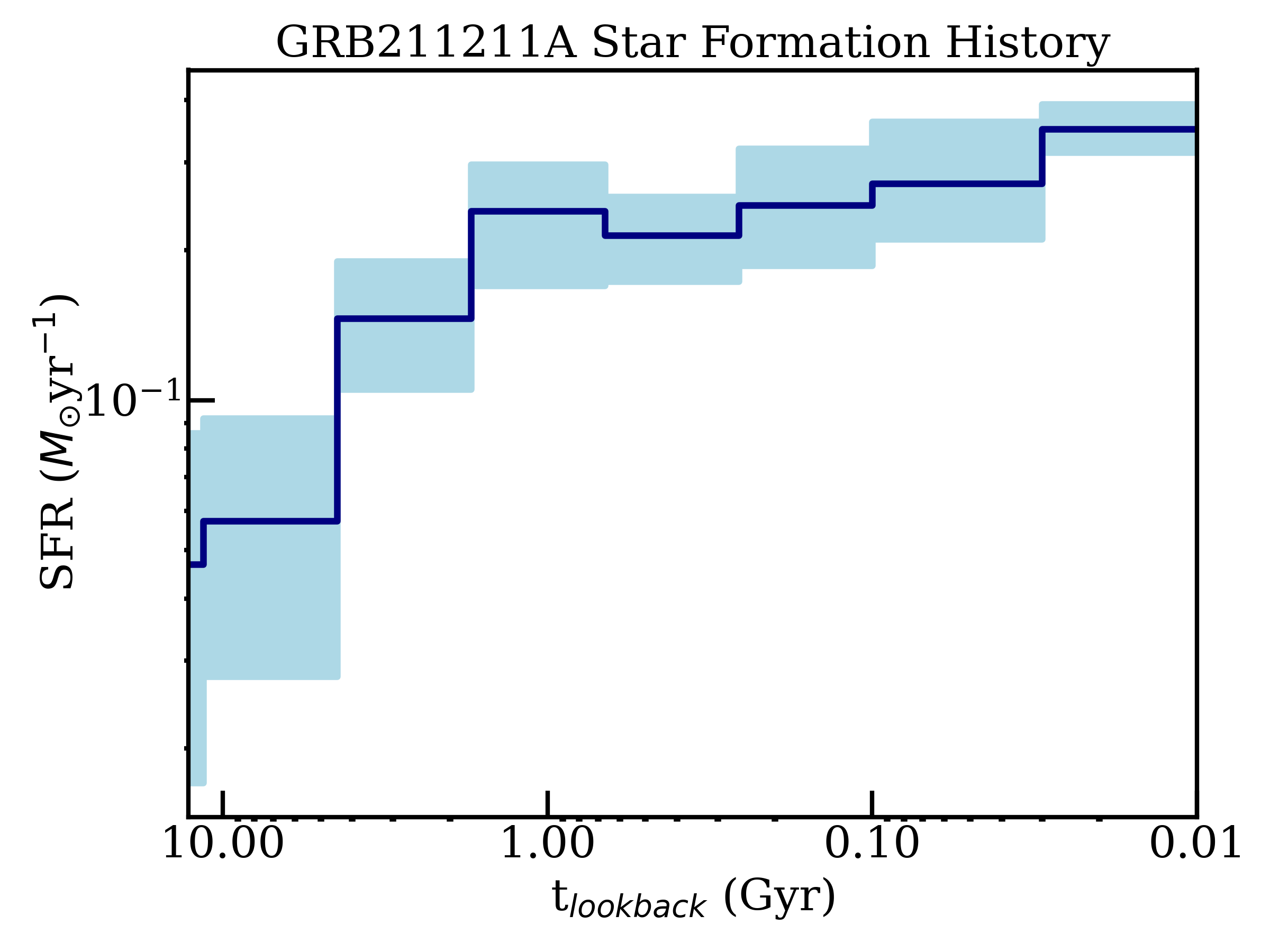}
\includegraphics[width=0.3\textwidth]{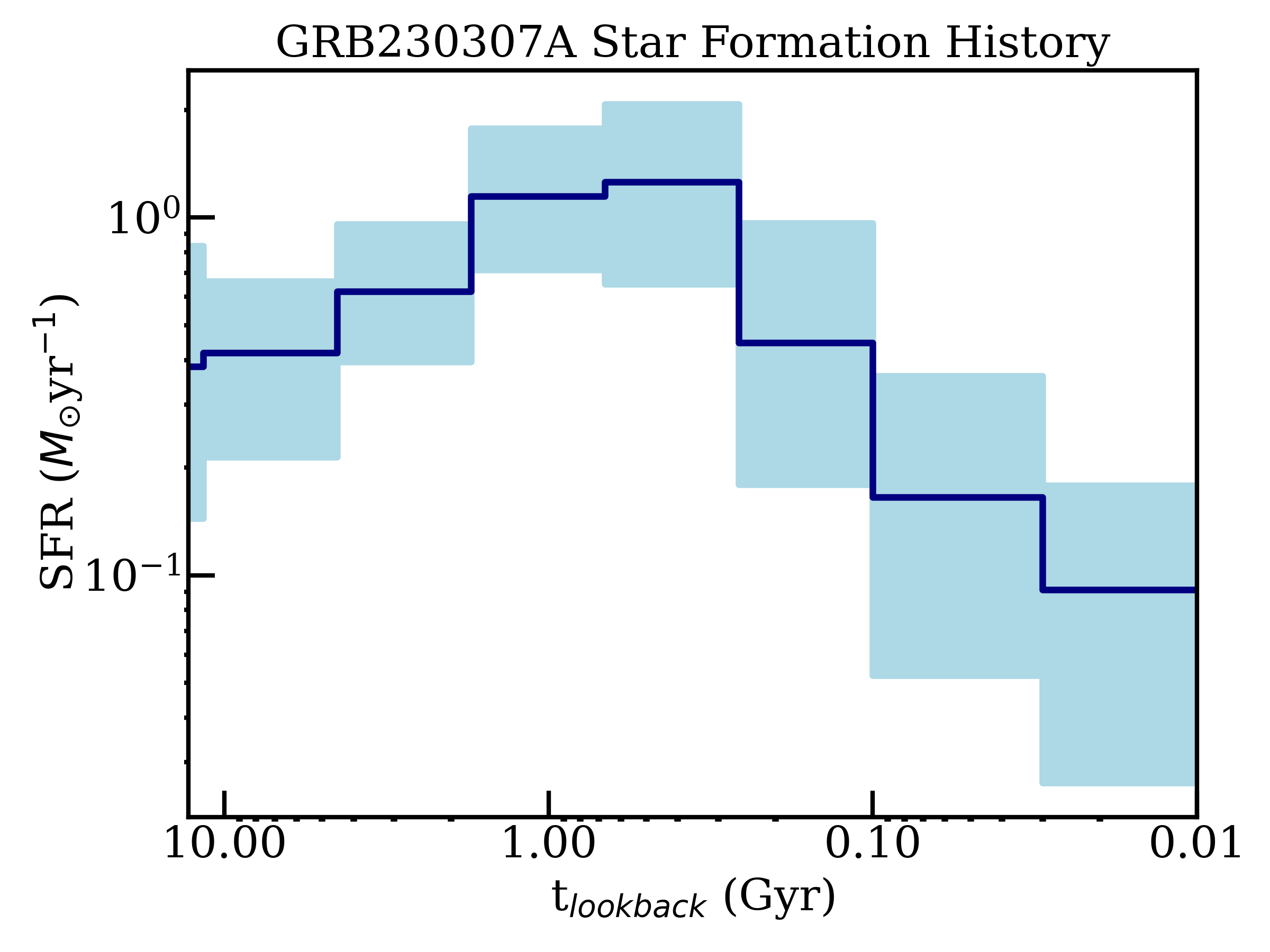}
\caption{The \texttt{Prospector}-derived star formation histories (SFHs) of the 12 GRB-KN host galaxies in our sample. This sample has a diverse array of SFHs and recent star formation bursts, hinting that some events might be more capable of enriching their environmental star-forming gas with $r$-process elements than others.
\label{fig:sfh}}
\end{figure*}

We obtained host associations for the short GRBs and long GRBs 060614 and 211211A in \citet{BRIGHT-I}. GW170817 is associated to host galaxy NGC 4993 \citep{bbf+17, llt+17, pht+17, kfb+2022}. \citet{lgs+2024} claimed the host of GRB 230307A is a galaxy at $z = 0.065$, $\approx 40$~kpc offset from the transient, given the tentative Te~III detection at the same redshift and a low probability of chance coincidence of $P_{cc} = 0.09$. Using the classification scheme from \citet{BRIGHT-I}, ten of these events have ``Gold" standard host associations ($P_{cc} < 0.02$), and two (GRBs 160821B and 230307A) have ``Silver" standard host associations ($0.02 < P_{cc} < 0.09$); thus, this sample all has fairly robust host associations.

We collect host galaxy optical and IR photometry for this sample from \citet{hwf+05, cmi+06, gcg+06, gfp+06, GALEX-phot, gfl+09, fbf10, lb10, fb13,dtr+14, aaa+15, fmc+2016,  bbf+17, flr+2021, BRIGHT-I}, and \citet{rgl+2022}, as displayed in Table A1 from \citet{BRIGHT-I}. For the host of GRB 230307A, we find NUV host photometry from the \textit{GALEX} survey \citep{GALEX-phot}, $ugriz$ optical photometry from the Sloan Digital Sky Survey (SDSS; \citealt{SDSS2020}), and mid-IR photometry from WISE \citep{WISE}. We correct all photometry for Galactic extinction in the direction of each GRB, using the \citet{sf11} dust maps. We furthermore collect the reduced, extinction-corrected host spectra on the Broadband Repository for Investigating Gamma-ray burst Host Traits (BRIGHT\footnote{\url{http://bright.ciera.northwestern.edu}}), which are available for all host galaxies except for those of GRB 070809, GW170817, and GRB 230307A.

In Table \ref{tab:prop}, we provide the redshift and host association confidence level for this sample. We note that this GRB sample has a lower median redshift ($z \approx 0.1615$) than that reported for the full short GRB sample ($z \approx 0.64$) in \citet{BRIGHT-I}, \citet{BRIGHT-II}, and \citet{otd+2022}. This is an expected observational bias, given that KNe become much more difficult to detect with increasing redshift. As the distance of the KN to the host galaxy will also inform the $r$-process enrichment timescale (see Section \ref{sec:timescale}), we further provide the projected galactocentric offsets for this sample. We find that the GRB-KNe sample has a median projected physical offset of $\approx 6.6$~kpc, within the median and 68\% confidence interval for the full sample of short GRBs ($\approx 7.7^{+20.9}_{-6.1}$~kpc; \citealt{BRIGHT-I}).

\subsection{Short GRB Sample}
Given that most of the general short GRB population also likely derives from NS mergers, it is probable that the majority of events also produce some $r$-process materials even if their KNe are not detectable. As previously mentioned, KNe are difficult to detect beyond $z = 0.3$ with ground-based detectors \citep{rkp+2021}, and the majority of short GRBs are observed over this redshift limit. Thus, under the premise that short GRBs will also produce $r$-process, we include the short GRB samples from \citet{BRIGHT-I} to increase our population size. This sample contains an additional 74 short GRBs with confident host associations and offsets. Host galaxy properties (including stellar mass, star formation rate, and either photometric or spectroscopic redshifts) are available for 59 hosts in \citet{BRIGHT-II} and stellar mass and spectroscopic or photometric redshift estimates for another 11 hosts are described in \citet{nfc+2024}.

\section{Host Galaxy and Halo Properties}
\label{sec:sp_model}
\subsection{Stellar Population Modeling \& Properties}
The $r$-process enrichment timescale will be heavily influenced by host galaxy properties, such as stellar mass (see Section~\ref{sec:timescale}). We further seek to understand if host galaxy properties affect the capacity for an environment to be enriched from a single NS merger event. To determine the host galaxy stellar masses and star formation histories (SFHs), we use the stellar population inference code \texttt{Prospector} \citep{Leja2019, jlc+2021}. We jointly fit the photometry and spectroscopy (when available) for each host galaxy and employ the nested sampling fitting routine \texttt{dynesty} \citep{Dynesty} to produce posterior distributions for the stellar population properties of interest. Internally, \texttt{Prospector} utilizes the \texttt{MIST} models \citep{MIST} and \texttt{MILES} spectral libraries \citep{MILES} through \texttt{FSPS} (Flexible Stellar Population Synthesis) and \texttt{python-FSPS} \citep{FSPS_2009, FSPS_2010} to produce model spectral energy distributions (SEDs). 

For all \texttt{Prospector} fits, we utilize the \citet{kroupaIMF} initial mass function (IMF), the \citet{KriekandConroy13} dust attenuation model, which determines an offset from the \citet{calzetti2000} attenuation curve and the ratio of light attenuated from old to young stellar populations, and the \citet{gcb+05} mass-metallicity relation to probe realistic stellar mass and stellar metallicity combinations. Although all redshifts are spectroscopically confirmed, we allow redshift to be a free parameter in our fits with a tight ($\pm 0.01$) prior range around the host redshift to mitigate any uncertainty in the redshift that may arise from varying data reduction methods. In addition, we include the \citet{DraineandLi07} IR dust emission model and sample the polycyclic aromatic hydrocarbon mass fraction ($q_\textrm{pah}$), given that five GRB hosts in this sample have mid-IR photometry. For the hosts with a spectrum, we model their spectral continua with a $12^\textrm{th}$ order Chebyshev polynomial, apply a spectral smoothing model to normalize the spectra to their respective photometry, and determine a gas-phase metallicity based on their spectral line strengths. For the hosts with only photometry, we assume a solar gas-phase metallicity. We further adopt a nebular marginalization template to marginalize over the observed emission lines. We also employ a spectral noise inflation model to ensure that the spectra are not overweighted in the fit in comparison to the photometry and a pixel outlier model to marginalize over noise in the spectra. 

Finally, we incorporate the non-parametric \texttt{continuity} SFH in \texttt{Prospector}. This contrasts with previous work that has modeled short GRB hosts with parametric delayed-$\tau$ SFHs (e.g., \citealt{BRIGHT-II} which modeled 69 short GRB hosts). A non-parametric SFH is advantageous for this study as we are capable of modeling recent bursts of star formation, and can therefore place firmer constraints on the fraction of newly forming stars that are capable of being enriched with $r$-process elements from our sample. This SFH is constructed by assuming a constant star formation rate (SFR) within an age bin. We use eight age bins: the first two range from 0-30~Myr and 30-100~Myr, and the final six are log-spaced from 100~Myr to the age of the Universe at each host's redshift. 

In Table \ref{tab:prop}, we report the median and 68\% confidence interval for the stellar mass ($M_*$) and present-day SFR (average SFR from 0-100~Myr) for each GRB host. We note that in nearly all cases, the stellar masses from the non-parametric SFH \texttt{Prospector} fits are consistent within the error bars or higher by $\approx 0.1-0.5$ dex than those reported in \citet{BRIGHT-II} and \citet{lgs+2024}, which both employed parametric delayed-$\tau$ SFH \texttt{Prospector} fits. This is a well-known discrepancy between non-parametric and parametric SFH models \citep{Leja2019}. The majority of SFRs also differ slightly from those in \citet{BRIGHT-II} and \citet{lgs+2024}, although with no systematic offset to higher or lower values. This change is also due to the non-parametric versus parametric model choice. In Figure~\ref{fig:sfh}, we present the SFHs for each host. This sample appears to have a diversity in SFH shapes, with four having rising SFH, two with constant SFHs, five with falling, and one with a burst of star formation. For our full sample, the stellar mass median and 68\% confidence interval is $\log(M_*/M_\odot) = 9.64^{+1.45}_{-0.69}$, while the SFR = $0.42^{+8.96}_{-0.38} M_\odot$~yr$^{-1}$. The large confidence intervals reflect the diversity in host galaxy properties for this sample. Both properties are consistent with those of the entire short GRB host population \citep{BRIGHT-II}, suggesting the GRB-KN sample should be fairly representative of the observed NS merger population.

\subsection{Halo Properties}
Since many short GRBs are observed to be far offset from their host galaxy centers, within their hosts CGM \citep{BRIGHT-I, otd+2022}, the $r$-process transport will be affected by their hosts' dark matter halos (see Section \ref{sec:timescale}). Thus, we define several halo properties. We calculate halo masses ($M_h$) using the redshift-dependent $M_*$-$M_h$ relation defined by \texttt{UniverseMachine} \citep{behroozi2019}, using the $M_*$ median value for each GRB host determined in our \texttt{Prospector} fits. We use Equation (2) in \citet{sff+2019} to find the virial radius ($r_\textrm{vir}$) of each host in kpc, which is given by:
\begin{equation}
r_\textrm{vir} = 260~\textrm{kpc} \times  \Big(\frac{M_h}{10^{12} M_\odot}\Big)^{1/3}.
\end{equation}
We define the virial velocity ($V_\textrm{vir}$) as:
\begin{equation}
    V_\textrm{vir} = \sqrt{G M_h/r_\textrm{vir}},
\end{equation}
and the virial temperature $T_\textrm{vir}$ as:
\begin{equation}
    T_\textrm{vir} = \frac{\mu m_p V_\textrm{vir}^2}{2k_B} \: \textrm{K},
\end{equation}
where $\mu = 0.59$ is the molecular mass for fully ionized primordial gas and $m_p$ is the mass of a proton. We list the $M_h$, $r_\textrm{vir}$,  $V_\textrm{vir}$, and $T_\textrm{vir}$ for each host galaxy in Table~\ref{tab:prop}.

\begin{figure}
\centering
\includegraphics[width=0.5\textwidth]{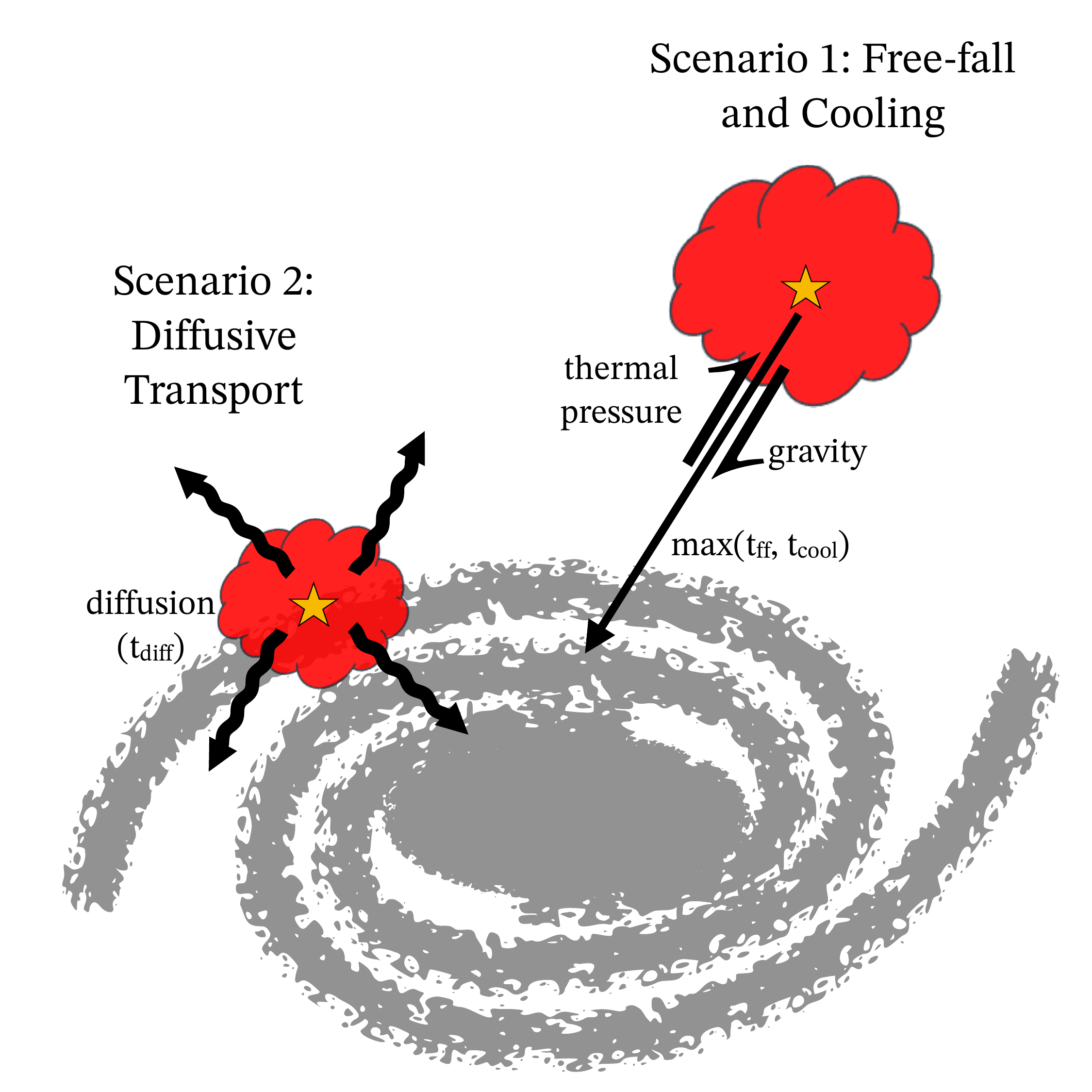}
\vspace{-0.1in}
\caption{The two scenarios we use to model the transport of $r$-process metals from the offset of the KNe (yellow stars) in our sample to the host galaxy center. In Scenario 1, we assume that the thermal pressure from hot halo gas can support the metals against gravitational free-fall if the gas is hot enough, and the timescale to transport the $r$-process will be the maximum of the free-fall ($t_\textrm{ff}$) and cooling ($t_\textrm{cool}$)  timescales. In Scenario 2, the metals are transported with turbulent diffusion mechanisms within the halo gas, and the fall back timescale is the diffusion timescale  ($t_\textrm{diff}$). The $r$-process enrichment timescale will be the faster of Scenario 1 or 2. As shown in Section \ref{sec:timescale}, diffusion is more efficient at transporting the metals at small KN offsets, whereas free-fall is more efficient at larger KN offsets, and cooling is only relevant for events highly offset from large halos.}
\label{fig:schematic}
\end{figure}

\section{$R$-Process Enrichment Timescale}
\label{sec:timescale}
To determine the total $r$-process enrichment timescale for each event, we consider two separate scenarios that can transport metals from the halo into the host interstellar medium (ISM): free-fall/cooling and diffusion. We show a schematic of these transport mechanisms in Figure \ref{fig:schematic}.  In the following subsections, we describe each of these scenarios and their impact on the $r$-process enrichment timescale.

\subsection{Scenario 1: Free-fall \& Cooling}
In Scenario 1, we consider free-fall and cooling. As the galaxy exerts a gravitational force on the $r$-process metals, thermal pressure from surrounding hot gas can keep the metals in a quasi-static equilibrium \citep[e.g.,][]{sff+2019}. If the cooling timescale ($t_\textrm{cool}$) is shorter than the timescale of gravitational free-fall ($t_\textrm{ff}$), suggesting the halo gas is not hot enough to support the metals against gravitational free-fall, the timescale for $r$-process metals to be transported to the center of the galaxy will be roughly $t_\textrm{ff}$ \citep{ro1977, silk1977, wr1978, sff+2019, sff+2020, fo2023}. Otherwise, the halo gas around the metals will need to cool, and the transport timescale will be $\approx t_\textrm{cool}$. We quantify the free-fall timescale as: $t_\textrm{ff} = \sqrt{2}~r_\textrm{KN}/v_c$ \citep[e.g.,][]{sff+2020}. Here, $r_\textrm{KN}$ is the deprojected KN galactocentric offset\footnote{Given that it is impossible to determine the $z$ direction of the KN, we assume the event lies at the median of an isotropic distribution of angles (30$^{\circ}$). Thus, to calculate the median deprojected offsets, we simply multiply observed projected offsets in Table \ref{tab:prop} by $\cos(30^{\circ})$, or $\sqrt{3}/2$.} and $v_c$ represents the circular velocity at $r_\textrm{KN}$, assuming a \citet{NFW} density profile with concentration $c$ from \citet{cws+2015}, which correlates $c$ with $z$ and $M_h$. Given the broad range of redshifts of the short GRB sample ($0 \lesssim z \lesssim 2.6$; \citealt{BRIGHT-I, BRIGHT-II}), we determine $c$ only at the median short GRB redshift ($z = 0.64$). We note that this assumption will add minimal uncertainty to our analysis.

Following the methods in \citet{sff+2020}, we characterize $t_\textrm{cool}$ as a function of $v_c$, the gas density ($\rho_\textrm{gas}$) and hydrogen gas density ($n_\textrm{H}$) surrounding the $r$-process metals at their initial location, and the \citet{wiersma2009} cooling function ($\Lambda_\textrm{T}$), which is dependent on $T_\textrm{vir}$ and the halo gas metallicity\footnote{We note that this metallicity is different from the gas-phase metallicity determined in our \texttt{Prospector} fits.}. Assuming $\rho_\textrm{gas} = 2.25 \mu m_p n_\textrm{H}$ for a completely ionized gas of primordial composition, we find that:
\begin{equation}
    t_\textrm{cool} = \frac{\rho_\textrm{gas} v_c^2}{n_\textrm{H}^2 ~ \Lambda_T(T_\textrm{vir}, Z)} \approx \frac{5.06 \mu^2 m_p^2 v_c^2}{\rho_\textrm{gas} ~ \Lambda_T(T_\textrm{vir}, Z)}.
\end{equation}
For simplicity, we assume that the gas metallicity is solar and that $\rho_\textrm{gas}$ follows an isothermal gas density profile, given by:
\begin{equation}
    \rho_\textrm{gas} = \frac{f_\textrm{CGM} f_b M_h}{4 \pi c r_s^2} \times \big(\frac{r_\textrm{KN}}{r_s}\big)^{-2},
\end{equation}
where $r_s = R_\textrm{vir}/c$ is the scale radius, $f_b = \Omega_b/\Omega_M$ is the baryon fraction, and $f_\textrm{CGM}$ is the fraction of baryons in the CGM. As described in \citet{sff+2020}, the true $f_\textrm{CGM}$ will depend on feedback and is related to the fraction of baryons that are kept in the halo versus the IGM. We adopt $f_\textrm{CGM} = 0.25$, based on median $f_\textrm{CGM}$ values determined for halos within the range $11 < \log(M_h/M_\odot) < 14$ from \texttt{EAGLE} simulations performed in \citet{odc+2020}. This $f_\textrm{CGM}$ selection only affects the events that occur highly offset ($\gtrsim 30$~kpc) from galaxies with larger halos ($\gtrsim 10^{12.5} M_\odot$), which is a small fraction of our observed GRB population.

We note that advection is another possible mechanism for metal transport  \citep[e.g.,][]{azh2022}, which we expect to occur on a similar timescale as the free-fall or cooling timescales. 

\begin{deluxetable*}{l|ccccc}
\tabletypesize{\normalsize}
\tablecolumns{10}
\tablewidth{0pc}
\tablecaption{Enrichment Timescales for GRB-KNe and GRB Populations
\label{tab:res}}
\tablehead{
\colhead{GRB} &
\colhead{Free-fall [Gyr]} &
\colhead{Cooling [Gyr]} &
\colhead{Diffusion [Gyr]} &
\colhead{Enrichment [Gyr]} &
\colhead{\% $M_*$ Enriched} 
}
\startdata
050709 & 0.094 & 3.0e-4 & 0.241 & Free-fall & 9-12\% \\ 
050724 & 0.043 & 1.6e-4 & 0.158 & Free-fall &  $<1$\% \\ 
060614 & 0.043 & 3.4e-6 & 0.028 & Diffusion & 44-57\% \\ 
070714 &  0.173 & 8.5e-3 & 1.169 & Free-fall & 54-71\% \\ 
070809  & 0.198 & 0.209 & 4.53 & Cooling & 4-45\% \\ 
130603B  &  0.105 & 8.8e-3 & 0.39 & Free-fall & 85-90 \% \\
150101B  & 0.067 & 3.0e-3 & 0.589 & Free-fall & 1\%  \\ 
160821B  & 0.226 & 1.4e-2 & 1.618 & Free-fall & $<1$\% \\ 
170817 &  0.047 & 7.6e-5 & 0.114 & Free-fall & $<1$\% \\ 
200522A & 0.041 & 5.3e-6 & 0.037 & Diffusion & 91-93\%  \\
211211A & 0.157 & 2.2e-2 & 0.649 & Free-fall & 17-20\%   \\ 
230307A &  0.449 & 0.129 & 5.396 & Free-fall & 1\% \\ \hline
GRB-KNe & -- & -- & -- & $0.100^{+0.114}_{-0.058}$ & 15\% \\
All GRBs & -- & -- & -- & $0.134^{+0.171}_{-0.083}$ & 59\% \\
\enddata
\tablecomments{The free-fall, cooling, diffusion and $r$-process enrichment timescales for our GRB-KN sample. The enrichment timescale is taken as the minimum of the diffusion timescale with the maximum of the free-fall and cooling timescales. We further present ranges (from uncertainties on the host SFR) of the percent of stellar mass enriched from the single GRB event after the $r$-process enrichment timescale to $z=0$ ($p_\textrm{enrich}$). Finally, we show the median and 68\% confidence interval of the enrichment timescale for the GRB-KN and entire GRB sample studied in this work, as well as the median $p_\textrm{enrich}$ for these samples.}
\end{deluxetable*}

\begin{figure*}
\centering
\includegraphics[width=1.0\textwidth]{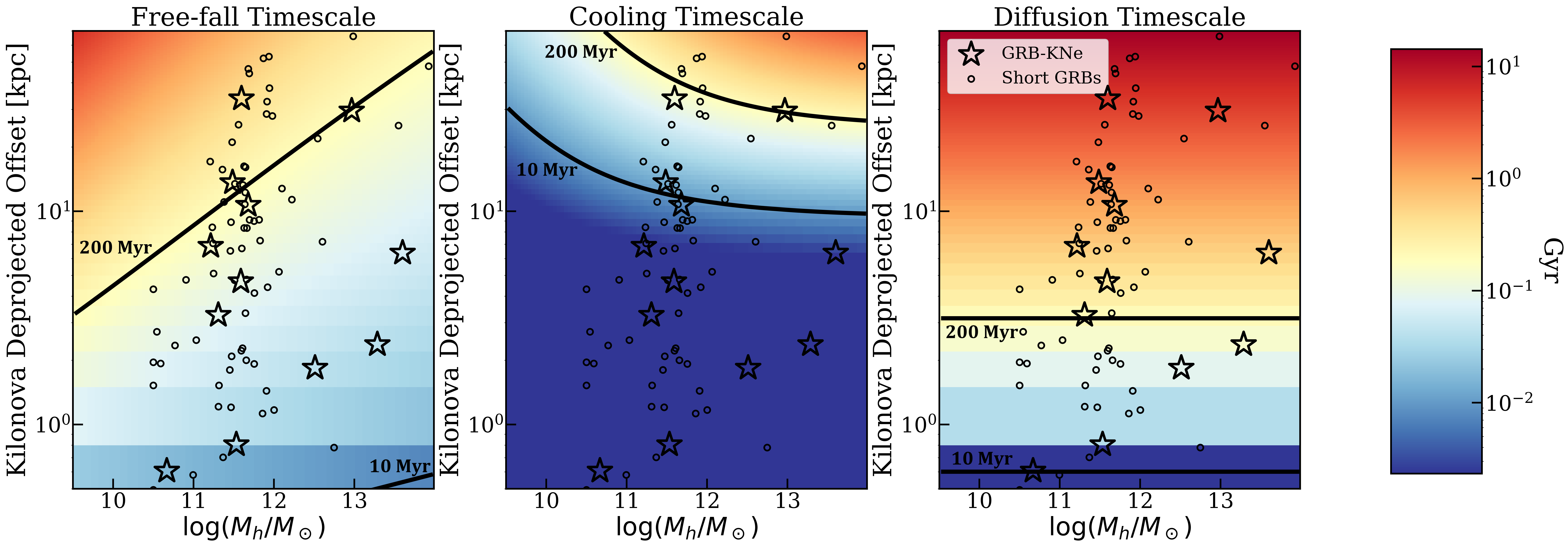}
\vspace{-0.1in}
\caption{Grids in KN deprojected offset (kpc) versus halo mass ($\log(M_h/M_\odot)$) space for the three timescales considered in this work: free-fall (left), cooling (middle) and diffusion (right). All timescales are in Gyr. Our GRB-KN sample (Table \ref{tab:prop}) are shown by the stars and the short GRB sample in \citet{BRIGHT-II} and \citet{nfc+2024} are shown by the black circles. The black contour lines represent where 10~Myr and 200~Myr fall for each of the two timescales. The majority of free-fall and diffusion timescales for the KN and short GRB samples are $>10$~Myr, while the cooling timescales are generally much shorter.}
\label{fig:all_timescales}
\end{figure*}

\subsection{Scenario 2: Diffusion}
In Scenario 2, we consider metal transport in turbulent gas, which we will model as a diffusion process \citep[e.g.,][]{es2004,se2004}. Turbulent diffusion is driven by physical processes like gas accretion, SN feedback, galactic winds, tidal disruption of satellite galaxies, and galaxy mergers, as well as other characteristics of the halo gas that might force diffusive transport \citep{es2004,ppj+2012}. Diffusion has often been used to model $r$-process transport in galaxies and generally helps explain observed $r$-process abundances and scatter \citep{mrn+2016, scr+2015, vqh+2015, bh2020, tyi2020, vpg+2020, vpb+2022, jsr+2023, krm+2023}. 

Here, we estimate the diffusion timescale using recent cosmological zoom-in simulations by Shah et al. (in prep.) of Milky Way-mass $10^{12} M_\odot$ halos from the Simulating the Universe with Refined Galaxy Environments (SURGE) project (\citealt{vbp+2021}; van de Voort et al. in prep.) based on the Auriga galaxy formation model \citep{grand2017}. In brief, Shah et al. (in prep.) study diffusive metal transport by injecting tracer dyes (passive scalars) into the CGM and quantifying the distance spread $S$, which encompasses the 68$^\textrm{th}$ percentile of the tracer dye. In these simulations, the rate of dye spread is approximately linear in time and can be modeled as
\begin{equation}
    \frac{\mathrm{d}S}{\mathrm{dt}} = K * \sigma_{\rm vel} [\textrm{km/s}] ^ b, 
\end{equation}
where $K = 0.55/(0.2~\textrm{Gyr})$ and $b=0.473$.
The diffusion rate depends on the gas velocity dispersion profile which follows the radial dependence
\begin{equation}
    \sigma_{\rm vel} (r) [{\rm km/s}] = A * r [{\rm kpc}]^{-B},    
\end{equation}
where $\sigma_{\rm vel}$ is evaluated at $\rm 1~kpc$ scales, $A=155$, and $B=0.7$.
The numerical coefficients for these relations are derived by averaging properties of three simulations with halo masses $10^{11} M_\odot$, $10^{12} M_\odot$, and $10^{13} M_\odot$. We did not find clear evidence for a halo mass dependence in these simulations, so for now we assume the metal diffusion rate is constant with halo mass, though an additional study is clearly warranted. The dependence of the diffusion timescale on $r_{\rm KN}$ can be found by solving:
\begin{equation}
    \int_{0}^{2r_\textrm{KN}}\frac{\textrm{d}S}{K*{\sigma_{\rm vel}(r_\textrm{KN}-S/2)}^b}= \int_{0}^{t_\textrm{diff}}\mathrm{dt}
\end{equation}
which assumes that diffusion is effective if $S$ surpasses $2r_\textrm{KN}$.
Solving for $t_\textrm{diff}$ gives the final result:
\begin{equation}
    t_\textrm{diff} [\textrm{Gyr}]= 0.05*r_\textrm{KN} [\textrm{kpc}]^{1.331}.
\end{equation}

\subsection{Timescale Results}
We present the free-fall, cooling (Scenario 1), and diffusion (Scenario 2) timescales for the GRB-KN sample in Table \ref{tab:res}.  We assume that the $r$-process enrichment timescale will be the more efficient process (or minimum timescale) between the two scenarios quantified in Figure~\ref{fig:schematic}; thus, the enrichment timescale is $\min(t_\textrm{diff}, \max(t_\textrm{ff}, t_\textrm{cool}))$. In Figure \ref{fig:all_timescales}, we compare the free-fall, cooling, and diffusion timescales in $r_\textrm{KN}$ versus halo mass. For the GRB-KNe sample, we find that free-fall timescales range from $\approx 40-450$~Myr, cooling timescales range from $\approx 0.003-130$~Myr, and diffusion timescales range from $\approx 30$~Myr - $5$~Gyr. When including the short GRB populations from \citet{BRIGHT-II} and \citet{nfc+2024}, we find a wider range of all three timescales: $t_\textrm{ff} \approx 20-600$~Myr, $t_\textrm{cool} \approx 700$~yr -$1.6$~Gyr, and $t_\textrm{diff} \approx 7$~Myr -$13$~Gyr. Generally, $t_\textrm{ff} > t_\textrm{cool}$, except at very large offsets from large halos, which suggests that cooling is not typically a rate limiting step in transporting the $r$-process. 

As we seek to understand how quickly an environment can be enriched with $r$-process from NS mergers, we compare these timescales to possible minimum NS merger delay times. From stellar population synthesis simulations of the Galactic binary NS (BNS) population, it has been estimated that minimum delay time is $\approx 10$~Myr \citep{bkv2002, dbk+2012, vns+18, bp2019}. A larger minimum delay time of $\approx 200$~Myr has been determined from the observations of the short GRB host population \citep{Zevin+DTD}, with typical delay time of $\approx 3-7$~Gyr \citep{Nakar2006, bfp+07, Jeong2010, Hao2013, Wanderman2015, Anand2018}, and Galactic BNS systems \citep{tkf+17,am19}. Given that the majority of free-fall and diffusion timescales for the full GRB-KNe and short GRB populations are $> 100$~Myr, this indicates that the enrichment timescale is about or more than the minimum NS merger delay time. Thus, these findings hint that the enrichment timescale is non-trivial in comparison to delay times, and that environments are not immediately enriched following an event.

\begin{figure}
\centering
\includegraphics[width=0.5\textwidth]{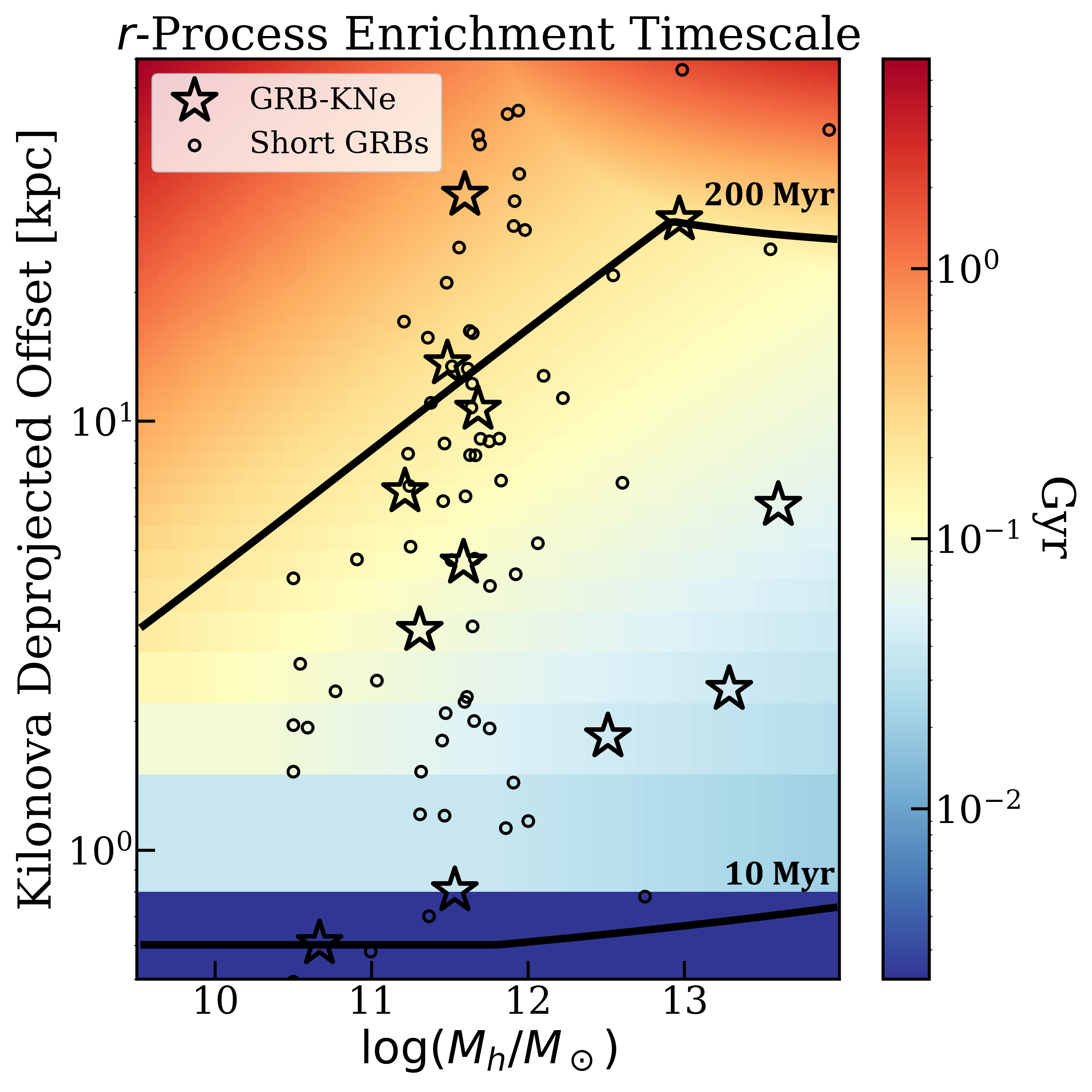}
\vspace{-0.1in}
\caption{The same as Figure \ref{fig:all_timescales}, but for the $r$-process enrichment timescale, which is the minimum of the cooling or diffusion timescale. We find that both the sample of KNe and short GRBs have significant enrichment timescales, on order of a $\approx 100$~Myr, compared to the minimum expected delay time for NS mergers ($\approx 10-200$~Myr).}
\label{fig:enrich_t}
\end{figure}

In Figure \ref{fig:enrich_t} and Table~\ref{tab:res}, we show the $r$-process enrichment timescales in the $r_\textrm{KN}$-halo mass parameter space along with those determined for the GRB-KN and short GRB samples. As expected, we find that at small deprojected offsets ($\lesssim 1$~kpc) and smaller halo masses ($M_h\lesssim 10^{11.55} M_\odot$), diffusion is more efficient at transporting the metals. Only two of the GRB-KNe in our sample are within the region where diffusion is more efficient: GRBs 060614 and 200522A, which have the smallest projected offsets in this sample. We further find that five of the short GRBs in \citet{nfc+2024}, which all occurred within central locations of dwarf galaxy hosts ($M_\star < 10^9 M_\odot$), also fall within this region. The majority of both the GRB-KNe (9 events) and the short GRB (62 events) samples fall within the region where gravitational free-fall is the most efficient process. For events in hosts with high halo masses ($M_h \gtrsim 10^{13} M_\odot$) and that have large deprojected offsets ($\gtrsim 25 $~kpc), which includes one GRB-KN (GRB 070809) and three short GRBs, their $r$-process falls back into the host on the cooling timescale.
  
For our GRB-KN sample, we find a range of enrichment timescales: from $\approx 28$-$449$~Myr, with median and 68\% confidence interval of $100_{-58}^{+114}$~Myr. For the full short GRB host sample (excluding the GRB-KN sample), we find a larger range of enrichment timescales ($\approx7$~Myr to $1.6$~Gyr) and a higher median ($138_{-78}^{+177}$~Myr). When combining samples, we find a median and 68\% confidence interval of $134_{-83}^{+171}$~Myr for the enrichment timescale. We further notice that the enrichment timescales exceed $1$~Gyr when the GRB is highly offset from its host galaxy ($\gtrsim 20-30$~kpc in deprojected space for $M_h < 10^{11} M_\odot$ and $\gtrsim 45$~kpc for $M_h < 10^{12} M_\odot$). This only accounts for 2 short GRBs.

In related work, \citet{azh2022} argued that the enrichment timescale should be substantially longer the the free-fall timescale, and, thus, the results found here. They showed that Rayleigh-Taylor instabilities will cause the $r$-process ejecta to mix in the halo gas clouds and Kelvin-Helmholtz instabilities will fragment the clouds and force the ejecta to completely mix in the halo gas. The authors speculate that the $r$-process will return to star-forming regions in the host through diffusion and advection, and estimate $\gtrsim1$~Gyr enrichment timescales for Milky Way-like halos. They assume diffusion timescales are driven by diffusion coefficients $D\approx1$~kpc$^2$~Gyr$^{-1}$, where $t_\textrm{diff}=r_\textrm{KN}^2/D$, typically used to describe diffusion in the galactic disk rather than the halo \citep{es2004, bh2020}, and that the advection timescale is $\sim M_\textrm{halo}/\textrm{SFR}$, predicted for hot-mode accretion. Our results suggest that the diffusion timescales should be much shorter than their estimates, and given the relatively small radii where kilonovae occur, we suggest that the advection timescales should be much shorter and described by $t_\textrm{ff}$ and $t_\textrm{cool}$. However, this disagreement in enrichment timescales emphasizes the importance of a more detailed simulation study in the future.

In summary, our results highlight that environmental enrichment is significantly delayed from the merger and that host properties and merger location dictate the length of the enrichment timescale.

\section{Discussion}
\label{sec:discussion}

\begin{figure*}
\centering
\includegraphics[width=0.8\textwidth]{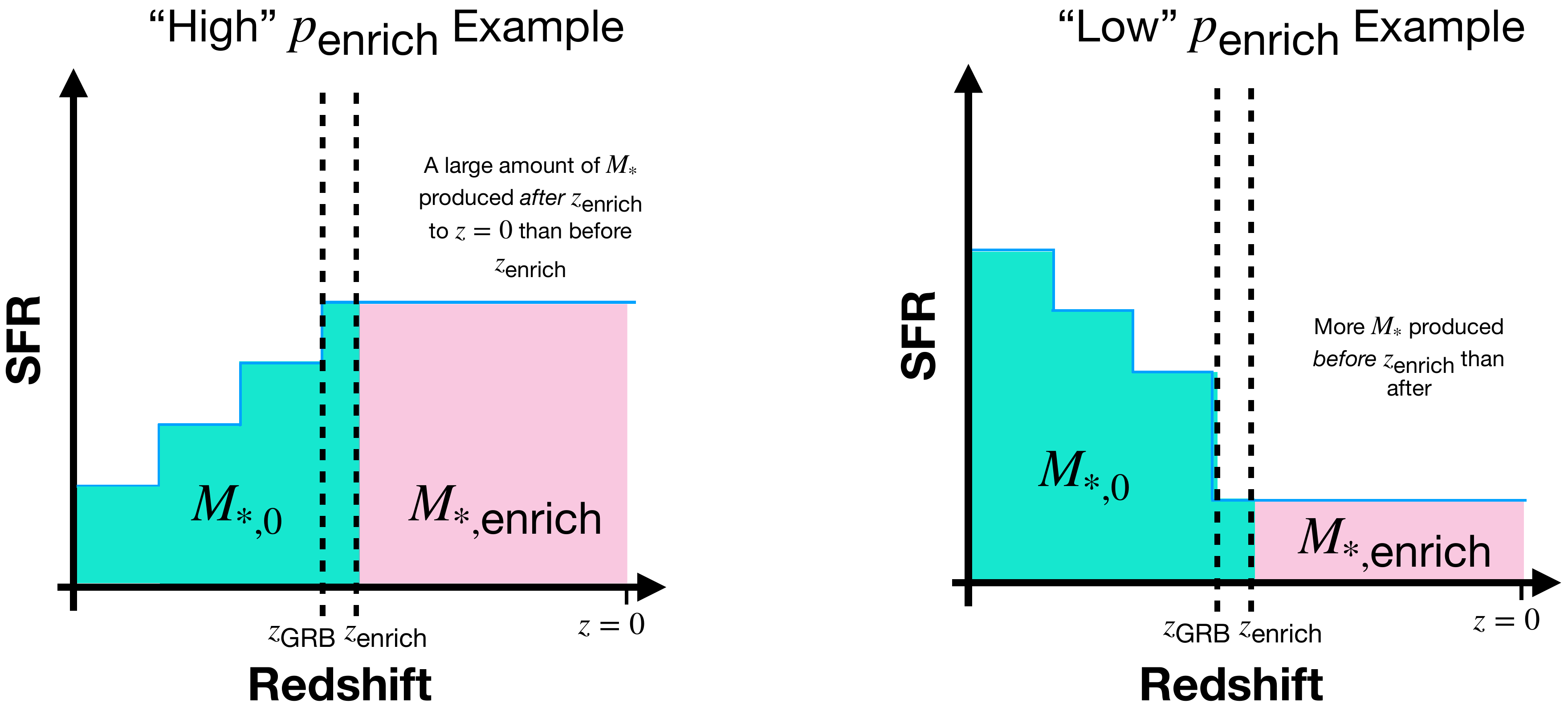}
\caption{An illustration of $p_\textrm{enrich}$ and how it is affected by the host's SFH. In the ``high" $p_\textrm{enrich}$ example, a substantial fraction of stellar mass is produced after $z_\textrm{enrich}$ in the host compared to the amount of stellar mass formed before $z_\textrm{enrich}$. In the ``low" $p_\textrm{enrich}$ example, the amount of stellar mass produced after $z_\textrm{enrich}$ is small in comparison the amount formed before.}
\label{fig:pe_illustrate}
\end{figure*}

\subsection{Enriched Stellar Mass}
Our next goal is to understand what host and NS merger properties dictate whether an NS merger actually enriches star-forming gas, or if the $r$-process is lost to the CGM or IGM. This, in turn, will help us determine whether the inferred $r$-process masses in, for example, Galactic metal poor stars and Local Group dwarf galaxies are directly comparable to the $r$-process masses produced from mergers. Thus, we perform a simple exercise to determine the percentage of the stellar mass of each host galaxy that is enriched by the observed GRB-KN event, projected forward in time (hereafter, $p_\textrm{enrich}$). We define $p_\textrm{enrich}$ as:
\begin{equation}
    p_\textrm{enrich} = \frac{M_{*,\textrm{enrich}}}{M_{*,0} + M_{*,\textrm{enrich}}},
\end{equation}
where $M_{*,0}$ is the stellar mass formed in the host before enrichment ($z_\textrm{enrich}$) and $M_{*,\textrm{enrich}}$ represents the stellar mass formed after $z_\textrm{enrich}$ to a future point in time. We assume that all stellar mass formed after $z_\textrm{enrich}$ will procure some $r$-process materials, and thus $M_{*,\textrm{enrich}}$ is the ``enriched" stellar mass. We show an illustration of $p_\textrm{enrich}$ in Figure~\ref{fig:pe_illustrate}. Note that $p_\textrm{enrich}$ characterizes the enrichment from a \textit{single} KN event, and fully describing $r$-process chemical evolution in galaxies requires accounting for other effects like the yield and rate of NS mergers over time, which are highly unconstrained both observationally and theoretically. It is estimated, for instance, that a typical Milky-Way sized galaxy has an $r$-process event rate of $\sim$tens Myr$^{-1}$, translating to tens of thousands of events per Gyr \citep{hbp+2018, rfb+2023}.

To determine $p_\textrm{enrich}$, we choose two lengths of time for a common comparison point within our sample: from $z_\textrm{enrich}$ to $z=0$, and 5~Gyr following $z_\textrm{enrich}$. We assume that the hosts have a constant SFR between $z_\textrm{GRB}$ to these future points, and calculate the amount of mass formed in the host over this time period. We note that this a reasonable assumption as it mandates that the hosts will have decreasing specific SFR (sSFR = SFR/$M_*$ in yr$^{-1}$) with growing stellar mass. A more sophisticated SFH extrapolation would require simulations, which goes beyond the scope of this work.  For the star-forming hosts in the sample, this extrapolation further also appears to keep the galaxies on the star-forming main sequence (SFMS; \citealt{Speagle2014, Whitaker2014, Leja2021}), a well-known galaxy correlation between a galaxies sSFR and stellar mass. We use the stellar mass-to-mass formed fraction, obtained from \texttt{Prospector}, to convert the mass formed into stellar mass and calculate $M_{*,\textrm{0}}$ and $M_{*,\textrm{enrich}}$, projected to $z=0$ and 5~Gyr after enrichment, for each host.

We ignore any effects of galactic outflows or winds, as well as the fact that some $r$-process may be ejected in the opposite direction of star-forming gas given the isotropic nature of KNe, that may prevent some of this stellar mass from being enriched. We use the 68\% confidence interval on their host SFRs (Table \ref{tab:prop}) to determine ranges of $p_\textrm{enrich}$ of each host at $z=0$ and 5 Gyr past enrichment. We also perform a similar calculation for the rest of the short GRB sample in \citet{BRIGHT-II} without detected KNe, using their reported stellar mass and SFR medians. With this analysis, we determine if the observed GRB population can be responsible for the stellar enrichment in their observed host galaxies.

\begin{figure}
\centering
\includegraphics[width=0.45\textwidth]{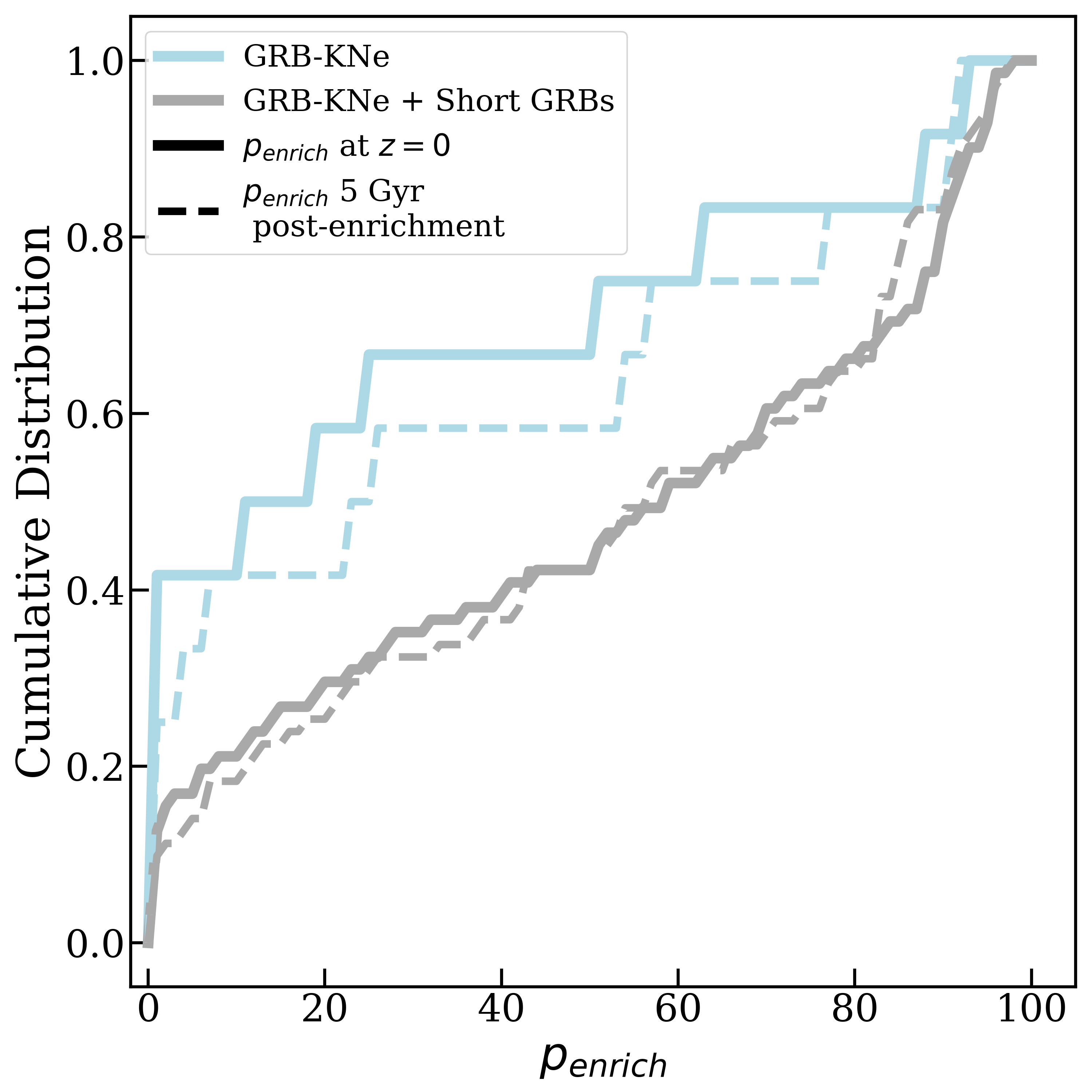}
\caption{CDFs of the percentage of stellar mass of the host galaxy that is enriched by the observed GRB-KN sample ($p_\textrm{enrich}$) at $z=0$ (solid blue) and by the full GRB-KN and short GRB sample (solid grey). We show the same for $p_\textrm{enrich}$ 5 Gyr after enrichment (dashed lines).  We find that the median $p_\textrm{enrich} = 13\%$ for the GRB-KN sample, suggesting that the majority of their $r$-process will be lost to the CGM or IGM. The median $p_\textrm{enrich}$ for the full sample is $59\%$. Given that there is a significant fraction of hosts that have $\lesssim 50\%$ enrichment, this suggests that not all environments are capable of being strongly enriched from a single event.}
\label{fig:pe_cdf}
\end{figure}

In Figure \ref{fig:pe_cdf}, we show cumulative distributions (CDF) of $p_\textrm{enrich}$ for the GRB-KN and full GRB samples for both the timescales between $z_\textrm{GRB}$ to $z=0$ and 5~Gyr after $z_\textrm{enrich}$. We first focus on $p_\textrm{enrich}$ determined for $z_\textrm{GRB}$ to $z=0$. We find that five of the GRB-KN hosts in our sample have $p_\textrm{enrich}\lesssim 1$\%, suggesting very little capacity for enrichment from a single $r$-process event, while three GRB-KN hosts have high $p_\textrm{enrich} \gtrsim 50$\%. Including the \citet{BRIGHT-II} short GRB population, we find that 13\% of the population has $p_\textrm{enrich}\lesssim 1$\% and $59\%$ of the population has high $p_\textrm{enrich} \gtrsim 50$\%. A key result is, therefore, that not all NS merger environments have the capacity to be significantly enriched with $r$-process material, despite the existence of an $r$-process source. This is especially true if the environment has already formed the majority of stellar mass before a merger event; thereby leaving a small fraction of newly-formed stellar mass for the NS merger to enrich (see Figure~\ref{fig:pe_illustrate}. Thus, we infer that a substantial fraction of $r$-process mass from NS mergers will be lost to the CGM or IGM, unlikely to ever be reincorporated in stars, implying that $r$-process ``losses" are indeed significant. Similar results were found in \citet{vpb+2022}, where stellar $r$-process abundances were measured from simulated NS mergers with and without natal kicks. When including natal kicks in there simulations, they find NS mergers are more likely to occur with larger galactocentric offsets and stellar $r$-process abundances in the host are decreased by $\approx 50\%$ than when not including kicks. This suggests a substantial amount of $r$-process is not being reincorporated into stars. We find that when we evolve all host galaxies for $5$~Gyr beyond $z_\textrm{enrich}$, $p_\textrm{enrich}$ changes only minimally with an average change of $5$\% for each GRB host. This implies that the length of time given to enrich star-forming gas has a small effect on the overall  $p_\textrm{enrich}$ from a single event.

We reiterate that our method for determining $p_\textrm{enrich}$ is simple and likely unsuitable for several specific cases. For instance, given that SFRs are generally higher at higher redshifts, an extrapolation of constant SFR to $z=0$ is unlikely to be realistic. Moreover, this method ignores any instances of possible future bursts of star formation after $z_\textrm{GRB}$. However, we emphasize that regardless of how $p_\textrm{enrich}$ is determined, it remains clear that some galaxies will not have the capacity to be enriched from mergers similar to the sample studied in this work, assuming that they do not have a future burst of star formation that would increase their capacity for enrichment.

\begin{figure*}
\centering
\includegraphics[width=1.0\textwidth]{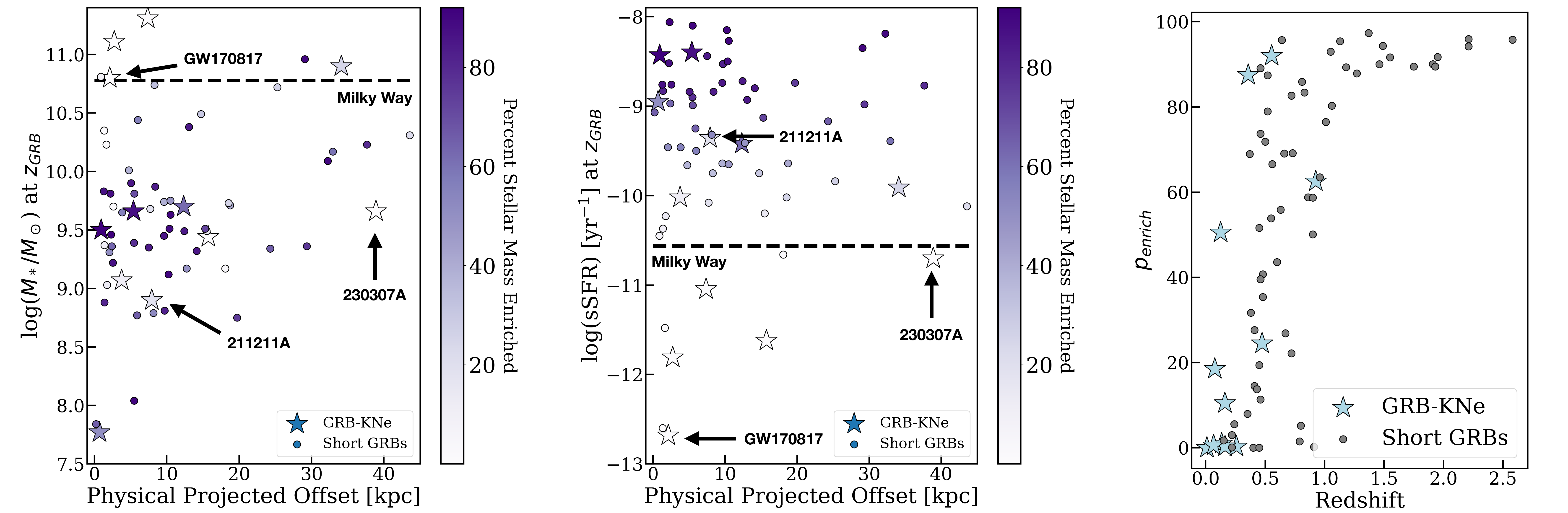}
\caption{\textit{Left:} The host galaxy stellar mass (at $z_\textrm{GRB}$) versus physical projected offset for our GRB-KN sample (stars) and the \citet{BRIGHT-II} short GRB sample (circles). The color indicates the percent of stellar mass enriched with $r$-process ($p_\textrm{enrich}$) from each event at $z=0$ and the black dashed lines denotes the Milky Way properties \citep{ln2015}. \textit{Middle:} The same figure, but for the specific SFR (SFR/$M_*$) versus physical projected offset. \textit{Right:} The redshift versus $p_\textrm{enrich}$ at $z=0$. While there does not appear to be a trend between $p_\textrm{enrich}$ and host stellar mass, GRB offset, or redshift (at $0.3 \lesssim z \lesssim 1$) we do find that $p_\textrm{enrich}$ increases with sSFR. Importantly, this emphasizes that NS mergers occurring in environments with little ongoing star formation will have no impact on their $r$-process enrichment. From this, we can infer that not all NS mergers are contributing to stellar $r$-process enrichment.}
\label{fig:perc_enrich}
\end{figure*}

\subsection{Trends with Host Properties and Offset}

To understand if there are specific factors that dictate how highly enriched an environment is from a single NS merger event, we compare $p_\textrm{enrich}$ (determined at $z=0$) to different host and GRB properties. Naively, we might expect that a highly offset event would have the majority of $r$-process lost to the CGM and IGM, resulting in a low  $p_\textrm{enrich}$. On the contrary, a higher mass galaxy with a larger population of stars to enrich might receive more enrichment from a single NS merger.

In Figure \ref{fig:perc_enrich}, we show a median $p_\textrm{enrich}$ in comparison to projected physical offset and host stellar mass for each GRB-KN and \citet{BRIGHT-II} short GRB sample. We find little correlation between $p_\textrm{enrich}$ and either of these properties, suggesting that the proximity to the host and its stellar mass has little or no effect on the host's capacity for enrichment. This is notable given that both properties strongly affect the enrichment timescale (Section~\ref{sec:timescale}). We do, however, observe that GRB hosts with $\log(M_*/M_\odot)\approx9.5$ have a higher degree of enrichment. This is likely explained by the fact that galaxies around this stellar mass have higher sSFRs, and are still building up their stellar mass.

Indeed, we see a much stronger trend between the sSFR at $z_\textrm{GRB}$ and $p_\textrm{enrich}$, as shown in Figure \ref{fig:perc_enrich}. Given the stronger dependence on sSFR rather than stellar mass or offset, this suggests that the enrichment timescale will only be a small perturbation in the host SFH. When considering all GRB-KNe and short GRBs with sSFR$>10^{-10}$~yr$^{-1}$ ($70\%$ of the population), all $p_\textrm{enrich}$ are $>19\%$, with median $p_\textrm{enrich}=79\%$. In contrast, the maximum $p_\textrm{enrich}$ for GRBs in hosts with low sSFR$<10^{-10}$~yr$^{-1}$ is $27\%$ with median $p_\textrm{enrich} = 2\%$. For very low values below sSFR$<10^{-10.5}$~yr$^{-1}$ ($18\%$ of the population), the maximum $p_\textrm{enrich}=5\%$, implying that environments with little to no ongoing star formation (and that have likely already built up the majority of their stellar mass) will not procure $r$-process material in newly-forming stars. Interestingly, we also find that the Milky Way lies within this sSFR regime \citep{ln2015}, with sSFR $\approx 10^{-10.6}$~yr$^{-1}$, implying that current or future Galactic NS mergers will have a minimal effect on the overall enrichment of the Milky Way. This is because the Milky Way has already formed most of its stars \citep{sneden2008}, so future NS mergers will impact a very low fraction of all Milky Way stars.

Overall, our results indicate that there are specific circumstances that allow an NS merger environment to be more enriched with $r$-process. Specifically, the amount of active star formation in the host will heavily influence the fraction of stellar mass enriched. This is intuitive, given that a galaxy needs ongoing star formation in order to be enriched by subsequent $r$-process events. For instance, the long delay-time NS merger GW170817, which occurred in an old, quiescent host galaxy, will have negligible effect on the $r$-process composition of the stars in its host. Future exploration using more complex simulations of galaxies and NS mergers is needed to better constrain the true rate of NS mergers that do indeed contribute $r$-process to their environments.

\subsection{Redshift Effects}

We further test if $p_\textrm{enrich}$ (determined at $z=0$) depends on the GRB redshift. It would be natural to assume, for instance, that the degree of enrichment might increase with GRB redshift as the $r$-process has more time to enrich newly forming stars. In Figure \ref{fig:perc_enrich}, we show $p_\textrm{enrich}$ versus redshift. We find that at $0.3 \lesssim z\lesssim1.0$ ($60\%$ of the full GRB sample), there is a large scatter in the values of $p_\textrm{enrich}$ and no obvious trend between $p_\textrm{enrich}$ and $z_\textrm{GRB}$. We verify this observed trend with a linear regression test, which indicates no linear correlation of $p_\textrm{enrich}$ within this redshift range. This indicates that the majority of the observed NS merger population will not have $p_\textrm{enrich}$ affected by redshift.

While there is no trend between $p_\textrm{enrich}$ and $z_\textrm{GRB}$ at $0.3 < z_\textrm{GRB} < 1$, we do find that at $z_\textrm{GRB} > 1$, all hosts are have high enrichment from a single event ($p_\textrm{enrich} > 76\%$) from a single event, while at $z < 0.3$, the majority of hosts have minimal enrichment ($p_\textrm{enrich} < 51\%$). Indeed, when including all GRBs, a linear regression test leads to a positive linear correlation between $z$ and $p_\textrm{enrich}$.

A natural subsequent question might be whether the majority of observed $r$-process in stars derives from high-redshift events. There are a couple of competing factors that make it difficult to answer this. For one, at higher redshifts, the sSFR of an average galaxy is higher, which would lead to more stars forming and a higher capacity for enrichment. However, the rates of NS mergers at higher redshifts are also lower than at lower redshifts, which is backed by the short GRB luminosity distribution \citep{Wanderman2015, lsb+16} and estimates on the NS merger delay time distribution \citep{bkv2002, Nakar2006, bfp+07, Jeong2010, dbk+2012, Hao2013, Wanderman2015, tkf+17, Anand2018, vns+18, am19, bp2019, Zevin+DTD}. Moreover, from both observational studies of tidally-disrupted dwarfs in the Milky Way \citep[e.g.,][]{njc+2022,ojf+2024} and simulation-based studies on Milky Way analogs \citep[e.g.,][]{vqh+2015, scr+2015, vpg+2020}, it is apparent that delayed $r$-process sources throughout galaxies star formation period are needed to explain their stars' chemical enrichment. Thus, it is unlikely that high redshift GRBs alone could be responsible for the bulk of observed $r$-process in stars.

\subsection{$r$-Process Enrichment at Low Metallicity}
Finally, we speculate on the $r$-process events that were responsible for the enrichment observed in Galactic metal poor stars and low metallicity Local Group dwarf galaxies. Enrichment at low metallicity prefers $r$-process from short merging delay times, which has been invoked many times \citep[e.g.,][]{astq2004, wpt2015}. If NS mergers are capable of merging as quickly as the minimum delay time predicted by theoretical work ($\approx 10$~Myr; \citealt{bkv2002, dbk+2012, vns+18, bp2019}), then enrichment at low metallicity is almost certainly created by NS mergers. However, if the delay time is longer, as predicted by observations of short GRBs and Galactic NS systems ($\approx 200$~Myr; \citealt{Nakar2006, bfp+07, Jeong2010, Hao2013, Wanderman2015, tkf+17, Anand2018, vns+18, am19, Zevin+DTD}), then NS mergers are unlikely to be a viable $r$-process source at low metallicity. Moreover, our work highlights that there is and additional timescale that needs to be incorporated when determining how quickly an environment is enriched after an $r$-process event: the enrichment timescale. Given that the enrichment timescale we find is not trivial ($\approx 134$~Myr), it becomes even more difficult for NS mergers to contribute at early times in low metallicity dwarf galaxies. Additionally, we show in Figure~\ref{fig:enrich_t} that enrichment timescales can be quite long for small, dwarf-galaxy sized halos ($\lesssim10^{11}M_\odot$), even at moderate NS merger offsets.

If collapsars and rare types of CCSNe were discovered to also produce $r$-process, this would be a more natural explanation for metal-poor stellar enrichment. Indeed, their delay times are much faster than NS mergers, $\approx10$~Myr, and they are typically found embedded in star-forming regions within their hosts \citep{kkp2008, prieto2008, bbf16}. Many works also suggest that a faster $r$-process channel like CCSNe is required to explain observed $r$-process abundances \citep{scr+2015, vqh+2015, ks2016, hbp+2018, cote2019, simonetti2019, ss2020}. However, there is still skepticism from both observations and theory that any class of CCSNe produces $r$-process; thus this may not be a valid explanation for enrichment at low metallicity. Furthermore, $r$-process material from CCSNe may be subject to other processes that prevent enrichment (like supernova winds) given their proximity to star-forming regions, or, contrarily, may lead to environments that are too well-enriched and thus not representative of the observed scatter in $r$-process abundances at low metallicity.

While we find that it would be unlikely for NS mergers to be responsible for enrichment in low metallicity environments, we caution the reader against assuming that it is \textit{impossible}. The delay time distribution of NS mergers is still unconstrained, especially at low metallicity. Thus, it is still ambiguous if NS systems can merge on more rapid timescales. Additionally, while uncommon within our sample, we do find that 9 GRBs (all with small physical deprojected offsets of $<1.5$~kpc) have enrichment timescales $<50$~Myr. Were these GRBs to also have similarly fast delay times, this may represent a population that could be capable of enriching low metallicity environments. An observational study of 11 short GRBs occurring in central locations within dwarf galaxy hosts \citep{nfc+2024} has further hinted that NS mergers may indeed be the source of $r$-process enrichment within low-mass galaxies. Thus, it is difficult to interpret at this time if another $r$-process channel is required to explain enrichment of all environments.

We finally note that the source of $r$-process enrichment at all metallicities remains an open, unanswered question. Several works have shown, using assumptions on the NS merger delay time distribution, that a more prompt channel than NS mergers, like collapsars, is also needed to explain the $r$-process abundance pattern of higher metallicity Milky Way disk stars (e.g., \citealt{cote2019, sbm2019}). Given that we find that the enrichment timescale is not trivial in comparison to the delay time, we suggest that it should be included into future chemical evolution simulations to better probe the dominant source of $r$-process in stars at the full metallicity range.

\section{Conclusion}
\label{sec:conclusion}
In this paper, we quantify the timescale for $r$-process from observed GRB-KNe and short GRBs to be transported back into their host and mix with the ISM. We further consider the impact of a single event on the stellar $r$-process enrichment of their host environments. We find that the typical $r$-process enrichment timescale is $134^{+171}_{-83}$~Myr. In comparison to minimum NS merger delay times of $\approx 10-200$~Myr, the enrichment timescale is significant, implying that there is an additional, substantial, delay following the merger to when the environment can be $r$-process enhanced. We find that only $\approx60\%$ of the observed GRB-KN and short GRB population will contribute significant $r$-process enrichment to the stellar mass of their host galaxy. The capacity of environmental $r$-process enrichment from a single event most strongly correlates with the amount of active star formation in the hosts, as hosts with little to no ongoing star formation have a low fraction of stellar mass enriched by a single event (including that of GW170817). This implies that a significant fraction of $r$-process mass from NS mergers either is not incorporated into a large fraction of stellar mass within their hosts or is lost to the CGM and IGM. 

Given our findings, it is useful to consider whether there is any possibility of detecting $r$-process lines in the IGM or CGM to observationally quantify the amount of ``lost" $r$-process mass. Unfortunately, we suspect that it is highly unlikely that $r$-process signatures can be detected outside of stars or stellar remnants. The high atomic mass of $r$-process elements implies that that material will be split into fewer atoms of each element and across many different isotopes. Moreover, $r$-process masses from single events tend to be quite small, suggesting that it will be extremely difficult to detect spectral emission or absorption features. Thus, future studies on $r$-process event rates should focus on the fraction of material making it back into star-forming gas and consider the impact of losses for any relevant observational comparison.

\section*{Acknowledgments}
We thank the anonymous referee for their comments and suggestions. We thank Claude-Andr{\'e} Faucher-Gigu{\`e}re, Alexa Gordon, and Michael Zevin for helpful discussions and commentary. The Fong Group at Northwestern acknowledges support by the National Science Foundation under grant Nos. AST-1909358, AST-2206494, AST-2308182, and CAREER grant No. AST-2047919. A.P.J. acknowledges support by the National Science Foundation under grants AST-2206264 and AST-2307599. W.F. gratefully acknowledges support by the David and Lucile Packard Foundation, the Alfred P. Sloan Foundation, and the Research Corporation for Science Advancement through Cottrell Scholar Award 28284. 
F.v.d.V. is supported by a Royal Society University Research Fellowship (URF$\backslash$R1$\backslash$191703).

This research was supported in part through the computational resources and staff contributions provided for the Quest high performance computing facility at Northwestern University which is jointly supported by the Office of the Provost, the Office for Research, and Northwestern University Information Technology.

\vspace{5mm}
\software{\texttt{Prospector} \citep{jlc+2021}, \texttt{Python-fsps} \citep{FSPS_2009, FSPS_2010}, \texttt{Dynesty} \citep{Dynesty}, \texttt{Astropy}}

\bibliography{refs}

\end{document}